\documentclass{article}
\usepackage{PRIMEarxiv}

\usepackage[utf8]{inputenc} 
\usepackage[T1]{fontenc}    
\usepackage{multirow}
\usepackage{xcolor}
\usepackage{longtable}
\usepackage{hyperref}       
\usepackage{url}            
\usepackage{booktabs}       
\usepackage{amsfonts}       
\usepackage{nicefrac}       
\usepackage{microtype}      
\usepackage{lipsum}
\usepackage{fancyhdr}       
\usepackage{graphicx}       
\graphicspath{{media/}}     
\DeclareUnicodeCharacter{0301}{\'{e}}
\title{Soft Skills Centrality in Graduate Studies Offerings
}

\author{
  María del Pilar García-Chitiva$^a$\href{https://orcid.org/0000-0001-6776-3422}{{\includegraphics[width=0.3cm]{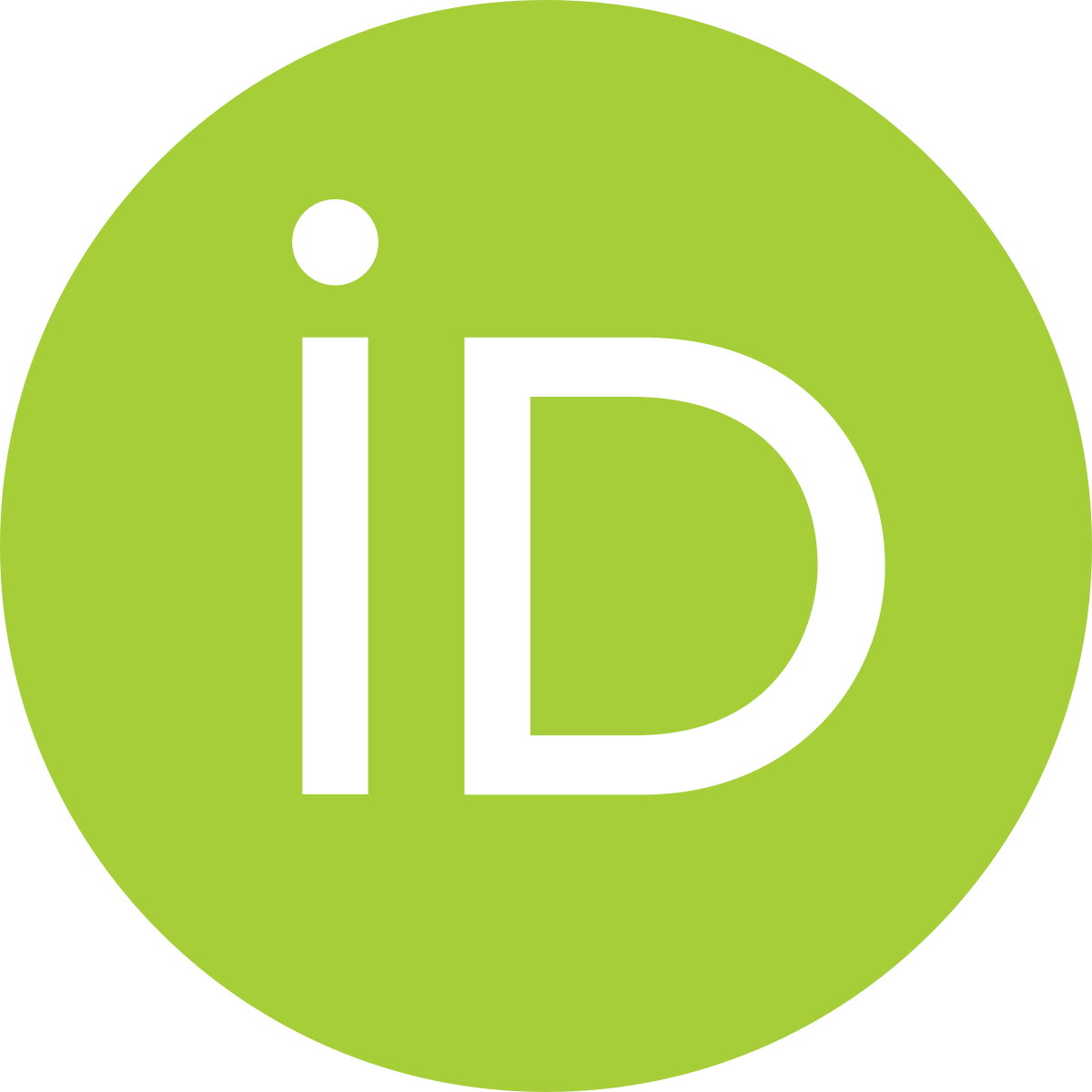}}} \& Juan C. Correa$^{a,b}$\href{https://orcid.org/0000-0002-0301-5641}{{\includegraphics[width=0.3cm]{orcid.png}}} \\
  $^a$ Instituto Tecnológico y de Estudios Superiores de Monterrey, Monterrey, México\\
  $^b$ Critical Centrality Institute, Monterrey, Mexico\\
  \texttt{pilargarciach@tec.mx, jcc@criticalcentrality.com} \\
}

\begin{document}
\maketitle

\begin{abstract}
Is it possible to measure how critical soft skills like leadership or teamwork are from the viewpoint of graduate studies offerings? This paper provides a conceptual and methodological framework that introduces the concept of a bipartite network as a practical way to estimate the importance of soft skills as socio-emotional abilities trained in graduate studies. We examined 230 graduate programs offered by 49 higher education institutions in Colombia to estimate the empirical importance of soft skills from the viewpoint of graduate studies offerings. The results show that: a) graduate programs in Colombia share 31 soft skills in their intended learning outcomes; b) the centrality of these skills varies as a function of the graduate pro- gram, although this variation was not statistically significant; and c)  while most central soft skills tend to be those related to creativity (i.e., creation or generation of ideas or projects), leadership (to lead or teamwork),  and analytical orientation (e.g., evaluating situations and solving problems), less central were those related to empathy (i.e., understanding others and acknowledgment of others), ethical thinking, and critical thinking, posing the question if too much emphasis on most visible skills might imply an unbalance in the opportunities to enhancing other soft skills such as ethical thinking.
\end{abstract}


\section{Introduction}

How well-trained are university students to face the challenges and difficulties of their working environment? This question has to do with UNESCO's global sustainability goals and public policies followed by each country to shape future education \cite{Jamali2023}. UNESCO's sustainable development goals provide an integral perspective on how professionals should be trained with enough disciplinary knowledge that allows them to provide some contributions to a series of global problems, including but not limited to poverty, hunger, or global warming \cite{Harvey2022}. As quality education is one of these UNESCO's goals, all educational institutions (from elementary to university) are summoned to cooperate, and this implies several challenges for Higher Education Institutions (HEIs) as their societal mission is to provide an education that transforms individuals into competent professionals contributing to their social and labor spheres \cite{brauer2021}. From this integral perspective, managers, headhunters, and members of corporate governance from several industry sectors often assume that professionals also provide their soft skills to exert and apply their professional knowledge to solve technical issues at work. As these skills are often mentioned in job offers and graduate programs intended to meet the needs of productive sectors, one might wonder how they are empirically taught in HEIs.

Soft skills training in higher education is not a novel topic. For example,   \cite{dell'aquila2016} pinpointed that competencies in higher education focus on distinguishing the so-called ``hard skills'' from those identified as ``soft skills.'' Hard skills are those explicitly related to professional knowledge in a discipline learned throughout a program (e.g., solving mathematical problems, doing statistical calculations, manipulating programming languages for specific purposes, or writing for scientific publication). In contrast, soft skills are related to intra and interpersonal characteristics, including but not limited to self-control, persistence,  leadership, or teamwork.

According to \cite{Scheerens2020}, soft skills are also \textit{social and emotional}. As a movement, the rise of soft skills has to do with keeping education responsive to societies' needs (e.g., changing demands of the labor market) and cultural trends (e.g., emotional intelligence to respect others' opinions and behavior). These needs and trends go hand-in-hand with developing social-emotional learning (SELs) programs for student-faculty interaction \cite{Awang-Hashim2022635}. This complex set of needs, demands, and trends co-exist in a labor market where graduates compete regarding employability, knowledge, skills, and experience \cite{succi2019,Succi2020}. The relevance of soft skills such as leadership, critical thinking, or persuasion is evident in several professional disciplines \cite{Coelho202278}, including but not limited to accountants \cite{Dolce2020}, data scientists \cite{Börner201812630}, mechanical engineers \cite{Rovida20231541}, physicians \cite{Riskiyana20222174}, librarians \cite{Hamid2022263}, and professionals of foreign languages \cite{Medvedeva2022} to mention just a few. Nonetheless, while these soft skills are paramount, recent evidence shows that students are not well-trained in how to talk about or show their soft skills in their resume. In a recent study, for example, it was reported that students tended to leave blank professional key sections in their LinkedIn profiles, could not communicate their unique value propositions, and described their experience section with poor messages, revealing that the profiles of already-employed individuals tended to be better than those unemployed \cite{Daniels202390}. Based on these facts, some scholars have claimed the use of LinkedIn as a pedagogical tool for careers and employability learning in higher education \cite{Healy2023106}.

Until this point, it should be evident that soft skills training is highly relevant for higher education, and their importance poses some questions that await institutional answers. For example, which factors should be considered vital when training soft skills in a graduate program and make this training different from those of undergraduate programs? How do students spot each university's institutional emphasis on soft skills training, and how should university staff design brochures and advertising material to reflect such emphasis to increase future students' attention and enrollment rates?

In an attempt to provide data-driven answers to these questions, a critical researcher must reach a sensitive balance between rigor and statistical representativeness when collecting empirical data. As the resources to reach a sample size statistically representative of all HEIs worldwide easily exceeds the capacity of any team,  the present study should be considered exploratory as it focuses on a representative sample of 230 graduate programs offered by 49 universities in Colombia.

We take the case of Colombia for two main reasons. First, this South American country has been a systematic object of study in higher education research during the last decade \cite{Alvarez2022,Duque2021669,Bradford2018909,Berry2014,Melguizo2011}. Despite these investigations, the literature on soft skills training remains scarce in Colombia and other Latin American nations \cite{Jaimes2022, Renteria2022}. The second reason is that this country joined the \textit{Organization for Economic Co-operation and Development} (OECD) on April 28, 2020, and this implies that some soft skills like leadership, teamwork, or creativity are paramount, as suggested by previous OECD reports \cite{Ocde2016-nq}. As these skills allow future professionals to work in a more globalized scenario where Colombia is expected to boost its current capacities \cite{Zarate2023}, this puts Colombia right in the middle of an ideal scenario to understand how the higher education system of a developing country responds to global expectations. In this context, for example, competent graduates might work on attracting foreign investors, but they require to master their persuasion skills should they want to be successful \cite{Dellavigna2010}. As graduates need to be trained to compete under the rules of new economic agents, it is widely known that executive business programs, for example, are a standard mechanism that facilitates the next generation of managers to succeed \cite{Lorange2019}. In this work, we also extend this view by sampling various graduate studies from other knowledge disciplines.

This work aims to introduce a conceptual and methodological helpful framework that illustrates the potential of combining bipartite network analysis with quantitative text analysis to estimate the relationship between soft skills and graduate programs. To reach our goal, the organization for the rest of this paper is as follows. First, we define soft skills and provide a literature review of their empirical evidence from the unique viewpoint of graduate studies. In other words, the present study does not contribute to the literature on soft skills in general. Instead, it does so to the literature focused on soft skills in graduate programs. Then, we describe the Colombian higher education system and the regulations that apply for creating and launching graduate studies in this country under two complementary local accreditation standards. Finally, we elaborate upon the centrality of soft skills from the perspective of complex networks \cite{Estrada2011}. Even though the concept of centrality is not novel in higher education research \cite{Glass2023}, our approach highlights \textit{bipartite networks} as a helpful way to quantify the relationship between graduate studies offerings and soft skills. An additional contribution of this work is visible in the section on materials and methods, where we describe the data collection and analysis procedure and provide their details following the standards of reproducible research \cite{Gandrud2018}. The writing of this work following reproducible research standards means that all interested readers can reproduce the results reported here, as both data sets and computational syntaxes are open and shared via a public repository and an appendix included after the references. In the end, the paper offers a discussion section that pinpoints relevant topics for further research.

\subsection{Soft Skills Conceptualization: A literature review from the viewpoint of Graduate Studies}

The concept of soft skills is polysemic \cite{bisquerra2007}. Its definition has different meanings like ``key competencies,'' ``transferable competencies,'' ``generic competencies,'' or ``socioemotional competencies.'' Furthermore, soft skills are considered integrated competencies, as they refer to the social interaction qualities that a person shows when solving problems, making decisions, and self-management while working with others \cite{dell'aquila2016}. As such, soft skills are capacities that let people adequately perform single and collective tasks \cite{vera2020}, which in turn work as a mechanism that people can leverage to get a job or a promotion through assertive and effective communication that shows an individual's critical thinking, teamwork, and valuable social interaction, involving both intra- and inter-personal abilities. Thus, soft skills are essential for companies' technology transfer and employee training \cite{Botke2018}.

Previous works show a connection between intra- and inter-personal abilities with socioemotional skills. For example, the renowned approach provided by \cite{goleman1998} relies on such a link. Other perspectives such as the one offered by \cite{zins2004}, focus on relevant soft skills training through the so-called ``socio-emotional learning'' (SEL) model. Due to the ample variety of conceptualizations toward soft skills, it is neither possible nor desirable to provide an exhaustive review of all these conceptualizations available in the literature. In contrast, our orientation focuses on a more nuanced conceptualization of soft skills in the context of higher education graduate programs. To reach this nuanced literature review, it becomes essential to understand the interplay between employers providing job opportunities and graduates looking for better jobs after completing a graduate program.

Most universities have a ``motto'' as a short phrase representing the institution's guiding principles, values, or aspirations. This motto serves as an ``added value'' or ``value promise'' for their graduate program offerings. Such ``promise'' or ``expectation'' communicates that their graduates have a valuable reputation due to the disciplinary competencies (hard skills) and soft skills such as critical thinking, decision-making, creativity, and teamwork they have learned throughout the official graduate program. These mechanisms work as reputational cues intended to help local employers to spot where the best generation of professionals comes from, so they can benefit by hiring them. However, the evidence reported in a series of empirical works shows gaps between the skills professionals should learn in university courses and the skills they are expected to exert in their jobs. This evidence seems to be transcultural as it has been reported from Malaysia \cite{abdullah2012}, Colombia and Mexico \cite{Berry2014}, the United States of America \cite{Börner201812630}, India \cite{sharvari2019}, Ukraine \cite{volkova2020}, Kosovo \cite{ziberi2021}, Italy \cite{Cattani2021,Dolce2020}, Poland \cite{cieciora2021}, Algeria \cite{Benlahcene2022}, and France \cite{Joie2023}. What follows shows a summary of soft skills from the unique viewpoint of HEIs and their graduate programs.

In Malaysia, \cite{abdullah2012} analyzed the existing gap between the soft skills training acquired by graduates of an electronic engineering program and the skills required by the labor sector as expressed by a sample of employers that included company directors, human resources directors, and production directors. Their results revealed that all participants perceived a mismatch for both hard and soft skills. For hard skills, they reported that professional weaknesses has to do with practical usage of the software tools, circuits construction, operate, troubleshoots systems and equipment, process, control and installation', as well as quality and reliability testing. For soft skills, they reported that professionals had weaknesses in information and communication technology skills, personal qualities, thinking skills, interpersonal skills, management skills, and general communication skills.

Even though the data collected through interviews with senior managers in six universities in Colombia and Mexico did not show a gap or mismatch between graduate students and their job opportunities, it showed a lack of bilingual communication skills for universities willing to increase internationalization activities that promote students interchange with allied universities in non-Spanish spoken countries \cite{Berry2014}. Such a circumstance contrasts with recent questionnaire-based research sampling engineers from Colombia. In that study, \cite{Jaimes2022}  identified the following ten soft skills as vital for the next generation of engineers as entrepreneurs: 1) responsibility, 2) integrity, 3) humility, 4) negotiation, 5) action coordination, 6) emotional competence, 7) leadership, 8) entrepreneurship, 9) critical thinking, and 10) sales performance.

In the United States of America, \cite{Börner201812630} used big data techniques to analyze more than 121 million jobs, more than 2 million university courses, and more than a million scientific publications between 2010 and 2016 to uncover the dynamic skill (mis-)alignment between academic push, industry pull, and educational offerings, paying special attention to the rapidly emerging areas of data science and data engineering. The authors also conducted a survey involving 20 labor market and educational domain experts from academia, industry, government, and the not-for-profit sector to examine the readability of the visualizations and the utility of results. Among their results, they reported that although HEIs aim to equip new generations of students with skills and expertise relevant to workforce participation for decades to come, their offerings sometimes misalign with commercial needs and new techniques forged at the frontiers of research. Once again, the study revealed the increasing importance of uniquely human skills, such as communication, negotiation, and persuasion, which are currently underexamined in research and undersupplied through education for the labor market, and suggested that in our data-driven economy, the demand for soft social skills, like teamwork and communication, increase with greater demand for hard technical skills and tools.

From India, \cite{sharvari2019} noted that although soft skills are critical and essential in the success of any professional, employers are complaining about the lack of soft skills training in university graduates, inhibiting them from gaining excellence in employability skills. To overcome this issue, \cite{sharvari2019} proposed that: ``the challenge is finding out the proper soft skills training and appropriate trainer who can impart training for Graduates,'' (p. 67) and such a challenge assumes that university authorities already know how employers perceive graduates.  \cite{sharvari2019} further suggested that employers perceive graduates with highly complex disciplinary competencies (hard skills) but need more soft skills for good job performance. They also established that the skills of self-motivation, creativity, teamwork, leadership, and identifying problems that arise in their work environment are the most desirable competencies for professionals in the labor market.

From Italy, \cite{Cattani2021} noted that the mentioned gaps are critical from the viewpoint of the assessment of labour market outcomes of HEIs. ``Students are also expected to either acquire or improve a broad set of horizontal competences, such as sophisticated relational and socio-emotional skills that are increasingly viewed as primary outcomes of higher education in a context of constantly evolving labour market'' (p. 2387). \cite{Cattani2021} classified a sample of graduate jobs on managerial and communication skills, to conduct an empirical analysis of the benefits of being employed in such occupations having completed a degree program. These beneﬁts were assessed across two moments between 2010 (when the economic crisis reached its peak in unemployment rates) and 2018 (when they collected more recent data concerning laboral force in Italy). Their results showed that only communication skills provided higher beneﬁts than discipline-related skills, and suggested that the Italian higher education system seems not to provide higher education graduates with distinctive managerial competences compared to non-graduate workers suggesting that these types of social skills can be developed through alternative learning paths, such as work-based experience or other extra-curricular activities.

\cite{volkova2020} examined students' perceptions of the importance of developing soft skills throughout graduate studies in a series of interviews addressed to 45 students of a graduate program at Alfred Nobel University in Ukraine. The results revealed a series of soft skills that proved to be systematically highlighted by interviewed students and grouped into the following three categories: 1) social and communication skills (communication skills, interpersonal skills, group work, leadership, social intelligence, responsibility, ethics of communication), 2)  cognitive skills (critical thinking, problem-solving skills, innovative thinking, intellectual workload management, self-study skills, information skills, time management), and 3) personal qualities and component of emotional intelligence (emotional intelligence, honesty, optimism, flexibility, creativity, motivation, and empathy). To develop such soft skills, university professors are also summoned to introduce a series of changes to their way of teaching, so they can include exercises to improve communication skills, cognitive skills, management skills, strategic skills, self-organization skills, and emotional competences.

From Poland, \cite{cieciora2021} proposed a system for collecting and analyzing reports on the employability of graduates. They used a sample of 31 graduates from a Polish university for its implementation. Their results showed that the Polish labor market highly values technical skills, especially computer skills, and business-related soft skills, such as teamwork. In addition, they established the need for HEIs to structure and offer curricula that promote the development of the required knowledge and soft skills coherently with the reality of the labor market. They suggested that developing these curricula should rely on a dialogue between the parties involved, HEIs, the labor market, and graduates.

From Algeria, \cite{Benlahcene2022} observed that the lack or ineffectiveness of ethics training and education inside and outside organizational settings has a detrimental impact on leaders’ ethical character. By conducting a series of semi-structured interviews with 15 leaders from public companies, they reported that public companies in Algeria suffer from several issues related to leaders’ ethics training and education. The findings also indicate that unethical leadership behaviors result from ineffective training programs and poor ethics education within public companies. Interestingly, these considerations, in turn, seem relevant for ambitious entrepreneurs looking for faster ways to increase the financial value of their start-ups \cite{Kuckertz2023}.

More recently, from France \cite{Joie2023} noted that an important goal for employees, companies, and HEIs is to find a relatively easy way to train soft skills that would allow transferability from the training program to specific work contexts in such a way that it has measurable effects on work outcomes. From this viewpoint, they analyzed the effects of a soft skills meta-cognition training program on self-efficacy and adaptive performance among large French company employees through a training course on interpersonal skills meta-cognition training. Their results revealed that identifying soft skills that one has and those that have not allow the person to develop personal strategies to enhance lacking competencies while strengthening those already developed, increasing self-efficacy and adaptive performance.

The professional profile individuals can get after completing their studies in a graduate program is a standard piece of information relevant to students, university authorities, and higher education policy-makers across nations. Although such information is available through advertising channels, all publicity efforts are expected to increase the number of enrolled students in all kinds of programs as much as possible to preserve the institutional sustainability of universities. From this perspective, it is often assumed that graduates can have a ``plus'' with symbolic connotations, such as ``entrepreneurs,'' ``leaders,'' or ``impactful'' characters. As these connotations are not directly associated with a subject's syllabus, program curriculum, or professional field, they are conceived as soft skills that are somehow trained during a program. Nonetheless, as evident from previous references, it is unclear how many soft skills should be trained in graduate programs, let alone the rationale for picking them as essential and how to train them. We regard this circumstance as an opportunity to uncover novel data-driven ways to tackle the problem of soft skills' importance for graduate programs. Given the relevant lessons that might emerge from an endeavor of this kind, we now switch the attention to the case of graduate program offerings in Colombia, assuming that this is just an initial step with the potential to stimulate similar endeavors across continents.

\subsection{Graduate Programs Offering in Colombia}

In Colombia, public and private institutions with or without profit-oriented purposes can freely propose graduate programs. The Minister of Education evaluates these proposals and authorizes their commercial launch, provided the institutional proponent complies with all requirements. The legal considerations for launching new graduate programs in this country are hierarchically organized as follows. The General Education Law (Law 115 from 1994) provides the broader reference, complemented by the public service of higher education (Law 30 from 1992), resources and competencies to organize educational services (Legislative act 1 from 2001), and a technical compendium (Decree 2566 from 2003, Regulatory Single Decree 1075 from 2015) with all educational norms that apply in this nation. Apart from these normative prescriptions, Decree 1330 from July 25, 2019, establishes the specific conditions for HEIs willing to launch new programs and evaluate the quality standards of these programs. By following these specific conditions, each institution is free to organize a technical committee in charge of documenting the specifics of a new graduate program proposal, along with the general description, the duration, the name of the courses, the list of professors, as well as the intended learning outcomes and the graduates' profile. 

In Colombia, higher education institutions offer three types of graduate programs (i.e., specializations, masters, and doctorates) with two types of accreditation that distinguish them. The first qualification is mandatory according to local regulations and is known as ``qualified accreditation'' (``\textit{registro calificado}''). The second is optional and is known as ``high-quality accreditation'' (``\textit{acreditado alta calidad}''). While the first qualification applies to all three types of graduate programs, high-quality accreditations apply for masters and doctorates exclusively, except for specializations in the health-related disciplines such as medicine, dentistry, epidemiology, or neurology. A practical way to differentiate specializations from masters and doctorates is the time required to attend classes to get a diploma. While the standard duration for specializations is one year or two semesters, masters require two years or four semesters. Doctorates, in contrast, can last at least three years or six semesters. In addition, developing research skills is essential for master's and doctorate but not for specializations. Thus, students enrolled in specializations do not have to prepare a thesis to get a diploma, but students enrolled in a master's or a doctorate need to work on a research thesis that shows how the student applied the scientific knowledge and the scientific method to understand any phenomenon with a critical and data-driven perspective.

The high-quality certification for graduate programs goes back to September 2008, with the Guidelines for High-quality accreditation for Master and Doctorate programs, issued by the National Accreditation Council (CNA) \cite{CNA2008}. The guidelines followed the standards of the National Council of Higher Education (CESU), an independent, related unit to the Minister of Education that is in charge of planning, assessing, coordinating, and recommending best practices for increasing the quality of higher education in Colombia. A critical reader might argue that, at this point, graduate studies offered in Colombia should show a significant change in how soft skills are visible for all professionals looking at continuing their studies. Although we are inclined to say that this visibility has undoubtedly increased since the Guidelines for High-quality accreditation publication, we are unaware of previous efforts supporting this idea. Section 6 of Decree 1330 from July 25, 2019, explicitly establishes the relevance of social abilities for professional education, claiming that all professional knowledge should be applied to their social environments following ethical principles and communicating technical advances of each discipline that benefit society which includes the individual development through the so-called long life learning. The contribution of the current work in this regard should be evident, as it is the first empirical attempt to uncover the explicit appearance of soft skills in the official description of graduate studies. The quantitative analysis of this information is possible through the concetps provided in the next section. 

\subsection{Soft Skills Centrality in Graduate Studies: A Bipartite Network Approach}

\begin{figure}[h!]
\begin{center}
\includegraphics[width=12cm]{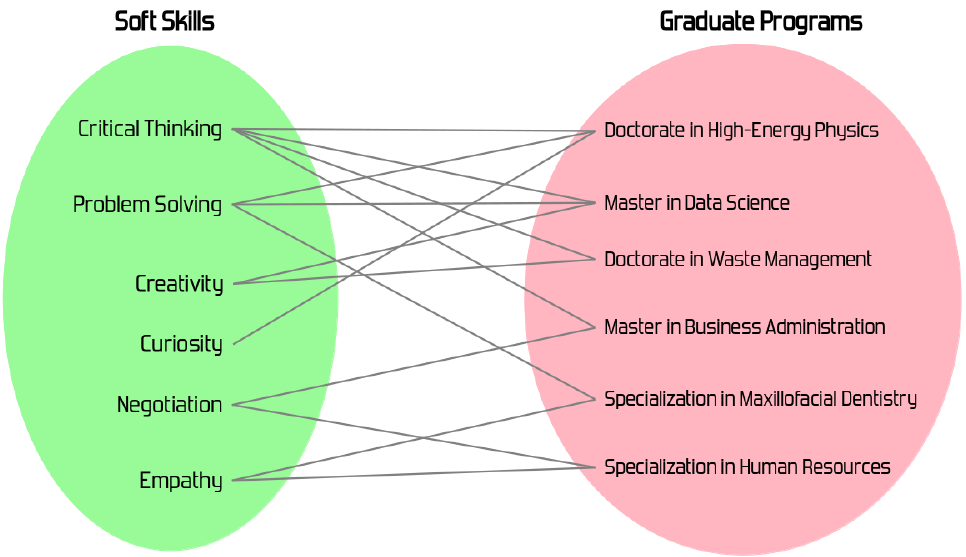}
\end{center}
\caption{A bipartite network visualization for soft skills centrality in graduate studies}
\label{F1}
\end{figure}

As mentioned earlier, social-emotional abilities mingle with other job requirements in non-trivial ways. For example, data scientists working with large volumes of customer reviews from an e-commerce platform can affect teamwork if big data technologies are not tested for automatically detecting patterns in natural language data through information-entropy-related metrics \cite{Correa2020}. In this case, the untested application of existing technologies introduces uncertainty or risk in achieving a project's goal, which might be intimidating for several professionals. These factors, however, should be solved in one way or another by professionals in charge of developing solutions. Soft skills in this case become a competitive advantage for the employees who work to satisfy customers' needs. Nonetheless, finding solutions might not be easy for graduates without field experience. From this viewpoint, the following questions frequently pop up among young graduates and employers: Are graduate studies appropriate for training professionals' soft skills? If so, how can we quantify the link between graduate studies offering and soft skills? Here, the concept of a bipartite network, as a particular case of complex network structures \cite{Estrada2011} emerges as a natural way to quantify the relationship between soft skills and graduate programs (see Figure \ref{F1}).

In a bipartite network, one is interested in understanding how two disjoint sets are connected. Figure \ref{F1} shows a hypothetical example with a non-exhaustive list of soft skills suggested by \cite{Scheerens2020} and a short and non-exhaustive list of graduate studies. The connection between these two sets relies on each link depicted as straight gray lines. In each set, each entity is formally known as a node and the  \textit{node adjacency} refers to the existence of an edge (i.e., straight gray line) that connects them. By looking at Figure \ref{F1}, it should be evident that the number of connections varies for each node in both sets. For example, critical thinking is the most connected soft skill as it connects with four programs (i.e., Doctorate in High-Energy Physics, Master in Data Science, Doctorate in Waste Management, and Master in Business Administration). Curiosity is the least connected as it only has one connection with one program (i.e., Doctorate in High-Energy Physics). Also, it should be evident that creativity, negotiation, and empathy have the same number of connections even though they connect with different programs. In addition, the bipartite network approach allows us to estimate the connectivity of each program. Thus, the doctorate in high-energy physics and the master in data science share the same number of connections (each one with three links), while the rest of the programs share the same number of connections (each one with two links).

\begin{figure}[h!]
\begin{center}
\includegraphics[width=14cm]{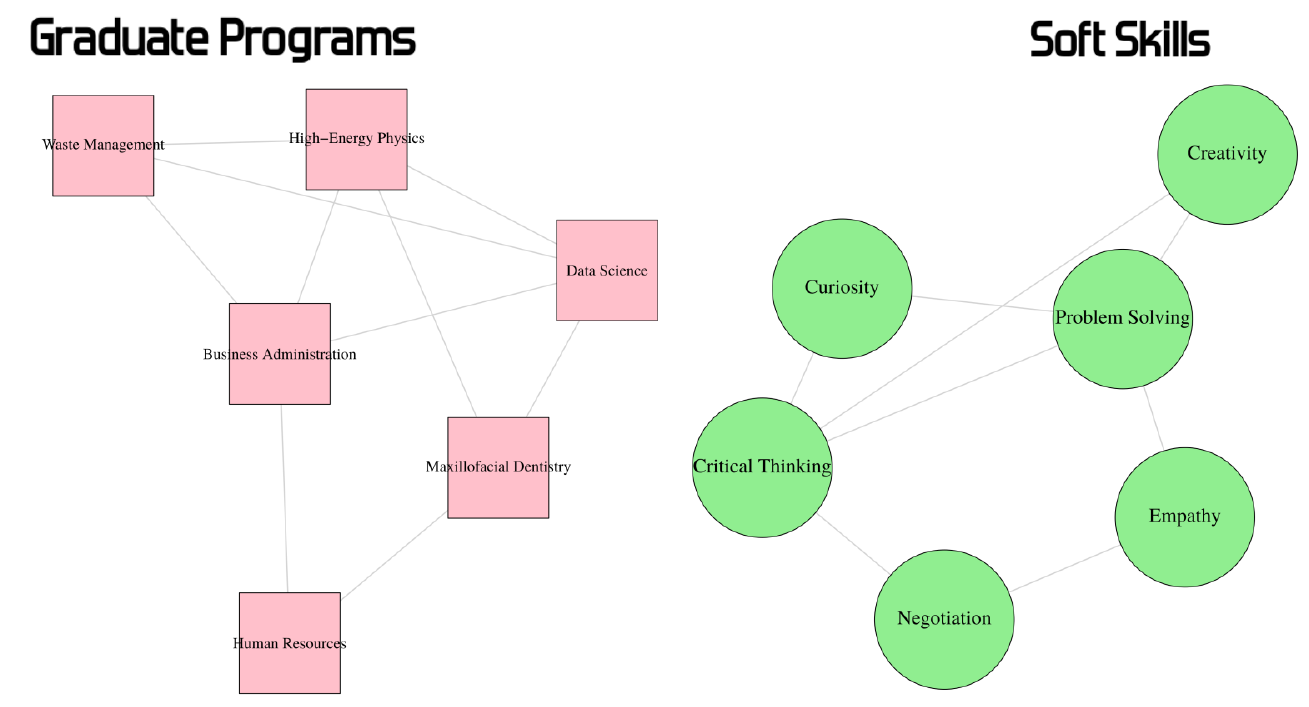}
\end{center}
\caption{Two resulting unipartite networks from the bipartite network visualization for soft skills and graduate studies}
\label{F2}
\end{figure}

Until this point, it should be clear that the concept of a bipartite network proves to be natural when it comes to understanding the relationship between two independent sets. With the bipartite network, one can examine the direct connections among the nodes of one set at a time (i.e., soft skills and graduate studies). To do this one can extract and visualize the unipartite projections of the original bipartite network. It is beyond the scope of this work to elaborate upon the mathematical and computational details related to these visualizations and projections. Still, interested readers are encouraged to revise the textbooks of \cite{Estrada2011} and \cite{Luke2015} to dive deep into these concepts. Figure \ref{F2} shows the one-mode or unipartite network projections for soft skills (with green nodes on the right) and graduate programs (with pink nodes on the left).

With these unipartite networks at hand, the calculus of \textit{node centrality} is fairly straightforward. However, a note of warning is worthy of mention here. Intuitively, node centrality refers to the network's ``most important'' node. One can think that counting the number of links or ties for each network's node and then ranking them from the most connected to the least connected might suffice to express this idea. \textit{Degree centrality} does just that. Nonetheless, in network analysis, several proxies exist for a node's importance. \cite{Oldham2019}, for example, summarized and analyzed 17 different metrics of node centrality. Despite this variety, it is important to understand the concept of centrality beyond mere intuition. According to \cite{Estrada2011}, ``a node is more central or more inﬂuential than another in a network if the degree of the ﬁrst is larger than that of the second'' (p. 122). Degree centrality, in other words, relates to a direct connection between any pair of nodes. Apart from this idea, one can also estimate how close each node is to all other nodes, which leads us to an alternative called \textit{closeness centrality}. \cite{Luke2015} defines closeness centrality as ``the inverse of the sum of all the distances between one node and all the other nodes in the network'' (p. 94). Another alternative is known as \textit{betweenness centrality}, which captures the extent that a node exists ``in-between'' pair of other nodes; that is, the edge between two nodes has to go through that node. A fourth alternative is known as \textit{Eigenvector centrality}, which measures nodes' transitive influence. A high eigenvector score means that a node is connected to many nodes with high scores. Although it is known that these centrality metrics share strong correlations in theoretical network models, their correlations for real-world networks tend to be lower \cite{ronqui2015,Oldham2019}. Given the statistical behavior of these metrics, it becomes relevant to analyze them in an unknown scenario like the one typically studied in higher education, in this case, the relevance of soft skills for graduate studies. 

To complete our bipartite network approach to the problem of quantifying soft skills centrality in graduate studies, it is mandatory to illustrate the principles that allow us to establish an objective connection or tie between any graduate program sampled from, say, the official website of a higher education institution, and any given soft skill. In essence, linking soft skills with graduate studies might be understood as one of the tasks of a behavioral data scientist. \cite{Saura2022} define behavioral data science as ``a new and emerging interdisciplinary field that combines techniques from behavioral sciences, psychology, sociology, economics, and business, and uses the processes from computer science, data-centric engineering, statistical models, information science, or mathematics, in order to understand and predict human behavior using artificial intelligence.'' (p. 2). 

One concrete way to define an objective link between soft skills and graduate programs is through the use of information retrieval techniques based on  \textit{natural language processing}. As per \cite{Manning2008}, information retrieval refers to ``finding material (usually documents) of an unstructured nature (usually text) that satisfies an information need from within large collections (usually stored on computers)'' (p. 1). This definition allows us to argue our fundamental claim in this work. The importance of soft skills training from the perspective of graduate programs can be tackled by harnessing the approach of bipartite networks as defined above. Such an approach remained unnoticed in higher education research. To the best of our knowledge, this is the first empirical study that opens new horizons to reveal the connection between soft skills and graduate programs offered by higher education institutions.

\begin{figure}[h!]
\begin{center}
\includegraphics[width=14cm]{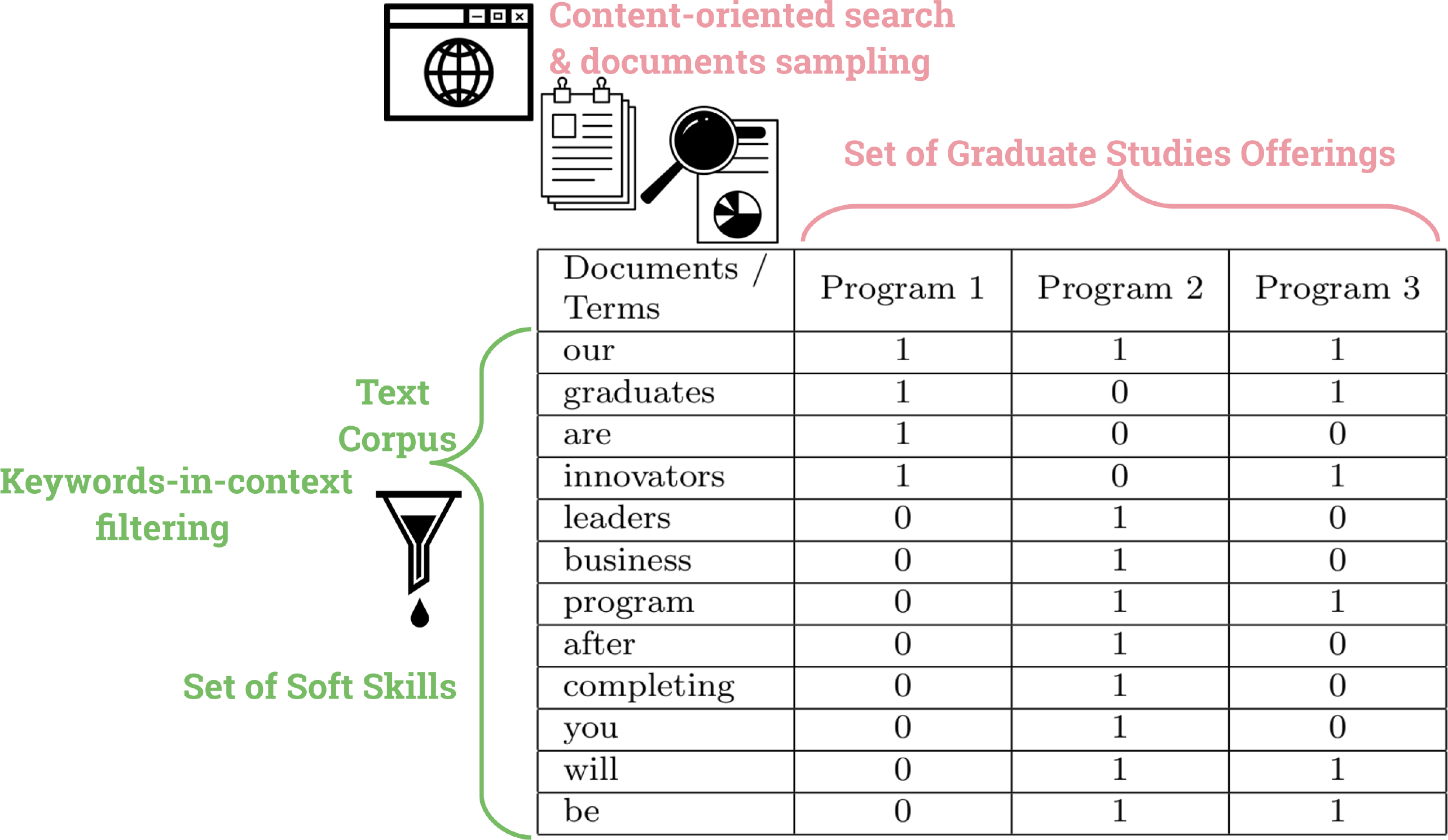}
\end{center}
\caption{A term-document matrix with cell entries that define objective links or ties between soft skills and graduate studies.}
\label{F3}
\end{figure}

An objective link between any program of graduate studies and any given soft skill can be defined through a term-document matrix. Figure \ref{F3} shows an example of a term-document matrix with three graduate studies as columns (i.e., Program 1, Program 2, Program 3) and a set of words as rows (e.g., innovators, leaders). In this matrix, cell entries can have one out of two possible values. If the word is explicitly present in the document, the value of the cell will be one and zero otherwise. 

This matrix is equivalent to an adjacency matrix, which is the standard input for bipartite network analyses.  In this matrix, the textual information from Programs 1, 2, and 3 can be read as follows: Program 1: ``\textit{our graduates are innovators},'' Program 2: ``\textit{after completing our program you will be business leaders},'' and Program 3: ``\textit{our graduates will be program innovators}.''  The example here shows that while Program 2 is designed for training business leaders, programs 1 and 3 are intended to train innovators. The apparent simplicity of this document-term matrix should not be regarded as a poor computational treatment, because other computational details are included in the process of estimating the link between soft skills and graduate programs (as will be evident in the section on materials and methods and in the appendix).

A critical reader who are not familiarized with the computational treatment to this collection of texts can easily find more details in the textbooks of \cite{Manning2008}, \cite{Estrada2011}, and \cite{Luke2015}. Nonetheless, our appendix provides helpful guidance for newcomers. Although this bipartite network approach has never been applied to the understanding of soft skills in the context of graduate programs, we take this as an opportunity to answer the following research questions:

\begin{itemize}
\item \textbf{Research Question 1}: Can we estimate the relationship between soft skills as socio-emotional abilities and graduate program offerings? And if so,
\item \textbf{Research Question 2}: Can we identify which soft skills tend to be more central in a sample of graduate programs? And if so,
\item \textbf{Research Question 3}: Are these soft skills equally tied to all accreditation standards and program types, or do they connect differently as a function of accreditation or program type?
\end{itemize}

\section{Materials and Method}

Our methodological approach is exploratory rather than confirmatory for two reasons. First, it provides evidence from a sample of 230 graduate programs offered by 49 universities in Colombia. The focus on this South American nation should be regarded as an initial step that stimulates future studies sampling more programs from different countries across continents. The second reason that makes a case for our exploratory endeavor is that it introduces the concept of a bipartite network to estimate the importance of soft skills from the textual information explicitly mentioned in graduate studies offerings. As such a conceptual approach has never been applied to higher education research, it is premature to claim that this study should be regarded as confirmatory.

Our methodological development elaborates upon the link between the theoretical understanding of soft skills as socio-emotional abilities developed by \cite{Scheerens2020} and a sample of graduate program offerings provided by a set of HEIs indexed by the Colombian Ministry of Education website and its National Information System for Higher Education (i.e., ``Sistema Nacional de Información de la Educación Superior'' or SNIES).

\subsection{Data sources}

By the end of February 2023, the National Information System for Higher Education reported 7,268 active graduate programs, and only 424 complied with conditions for high-quality accreditation. We used this information to conduct a query-based search within SNIES' official website with the option of ``public consultation by programs'' (i.e., ``consultas públicas por programas'') including the following additional filters: \texttt{Institution State}: ``Active''; \texttt{Branch type}: ``Principal''; \texttt{Program State}: ``Active''; \texttt{Academic Sector}: ``All'', and \texttt{Academic level}: ``Graduate studies''. This query-based search allowed us to retrieve a list with the names of all graduate studies programs during the first semester of 2022. 

The resulting list was downloaded as a Microsoft Excel file for pre-processing data, including a stratified sampling procedure with its selected programs. We used the names of the institutions and their programs to conduct a website-oriented search to identify and store the textual description of each selected program as captured by their online commercial offerings. After carefully exploring the official websites of sampled institutions, we identified and downloaded the official description of 282 graduate programs. Preliminary analyses of the information revealed that Colombian institutions do not follow a standard template or electronic format (e.g., docx, txt, html, or pdf) when documenting their programs and making them visible through their official websites. Given this variability, we also relied on historical reports from SNIES to ensure that our sampled programs had at least one promotion of graduates between 2018 and 2021, and we found that during this period, the total of graduates from private institutions (314,606) outnumbered that from public institutions (123,984). For all practical purposes, however, this work focuses on 230 valid programs (i.e., we discarded 52 programs because they had no graduates by December 2022, which was our temporal limit for sampling a current set of graduate programs). The distribution of sampled programs is summarized in Table \ref{T2}. 

\begin{table}[h!]
\centering
\caption{Random Stratified Sample of Graduate Studies during the first semester of 2022 in Colombia by Institution, Program, and Accreditation type}
\label{T2}
\begin{tabular}{|ll|c|c|c|}
\hline
\multicolumn{1}{|c|}{\begin{tabular}[c]{@{}c@{}}Accreditation \\ Type\end{tabular}}                         & Program Type   & \multicolumn{1}{l|}{Public} & \multicolumn{1}{l|}{Private} & \multicolumn{1}{l|}{Total} \\ \hline
\multicolumn{1}{|l|}{\multirow{3}{*}{\begin{tabular}[c]{@{}l@{}}High-quality\\ Accreditation\end{tabular}}} & Specialization & 3                           & 3                            & 6                          \\ \cline{2-5} 
\multicolumn{1}{|l|}{}                                                                                      & Masters        & 3                           & 8                            & 11                         \\ \cline{2-5} 
\multicolumn{1}{|l|}{}                                                                                      & Doctorate      & 5                           & 2                            & 7                          \\ \hline
\multicolumn{1}{|l|}{\multirow{3}{*}{\begin{tabular}[c]{@{}l@{}}Qualified\\ Accreditation\end{tabular}}}    & Specialization & 26                          & 83                           & 109                        \\ \cline{2-5} 
\multicolumn{1}{|l|}{}                                                                                      & Masters        & 32                          & 45                           & 77                         \\ \cline{2-5} 
\multicolumn{1}{|l|}{}                                                                                      & Doctorate      & 8                           & 12                           & 20                         \\ \hline
\multicolumn{2}{|c|}{Total}                                                                                                  & 77                          & 153                          & 230                        \\ \hline
\end{tabular}
\end{table}

\subsection{Data Analysis}
We conducted our analyses in the R system \cite{RCore2022}. The official records gathered from the Colombian Ministry of Education and its National Information System for Higher Education were processed with standard data science techniques \cite{Wickham2019}. Text analyses were treated with quantitative text-mining procedures \cite{Feinerer2008,Benoit2018}. 

Text analyses proceeded as follows. We created a local working folder directory that stores 230 files with the textual information from each graduate program. During the process of information gathering, we noticed that all programs were available as online websites (.html), and some of them did not have a downloadable version or had as portable document formats (.pdf) or Microsoft Word documents (.docx). We stored these documents in our local folder and used them to create a text corpus in Spanish. This corpus served as the input for a standard document-term matrix where documents were arranged as rows, words were arranged as columns, and the presence of a particular word in a specific document was automatically registered as one (1) and with zero (0) otherwise. 

Preliminary analyses on this document-term matrix allowed us to develop a double-step procedure to quantify the relationship between soft skills defined by \cite{Scheerens2020} and \cite{zins2004} and the information described in the academic program where these skills were explicitly mentioned. As a first step, we applied a \textit{keyword-in-context} search guided by a list of 50 terms and bigrams that matched the conceptual classification of socio-emotional skills proposed by \cite{Scheerens2020} and \cite{zins2004}. The \textit{keyword-in-context} search led us to build one detailed data frame for each keyword, where each keyword was stored along with the set of terms that explicitly appeared before and after the keyword in the corpus. We iterated this procedure and identified 43 keywords explicitly mentioned in at least one sampled program. We merged all resulting data frames into one single data set used as the input for our second step. We removed the words before and after the explicit keyword in our corpus and reduced the data frame to a filtered edge list. With the help of the \texttt{quanteda} R package \cite{Benoit2018}, we developed an ultimate term-document matrix with Spanish as the default input language. In this matrix, all numeric characters, punctuation characters, Spanish stopwords (i.e., most frequent words, including pronouns, prepositions, articles, and other part-of-the-speech tokens with nil semantic information), and common words that appeared in the educational jargon were removed to decrease matrix sparsity. This matrix was then coerced or transformed as a filtered edge list. According to \cite{Luke2015}, an edge list is a data format that depicts network information
by simply listing every tie in the network. In our case, a tie is just the connection between the keyword and a program. For example, if in ``\textit{Program A} the keyword ``\textit{lead}'' explicitly occurs as ``\textit{The graduate will be able to lead teams to achieve common institutional goals},'' here we have a tie between program \textit{A} and keyword \textit{lead}. This procedure allowed us to describe and quantify the explicit presence of these skills in our sampled documents by adopting a \textit{bipartite network analysis} \cite{Estrada2011}. In a bipartite network, also known as an affiliation or two-mode network, two types of nodes (i.e., soft skills and graduate programs) co-exist, and its co-existence is quantified through the number of times each node of one type is connected to nodes of the other types. Even though such analysis has been applied in social sciences \cite{Bail2016}, its application in higher education research remains ignored.

In this regard, the contribution of the present article is evident as it provides a helpful methodological approach for higher education researchers. This approach is developed following the standards of reproducible research with data and code available for those interested in knowing the computational details of data analysis \cite{Gandrud2018}.  The data, codes, and analyses are available in the Appendix of this article as well as in a public Github repository (\url{https://github.com/jcorrean/SoftSkillsUniversityPrograms}).

\section{Results}

\begin{figure}[h!]
\begin{center}
\includegraphics[width=1\textwidth]{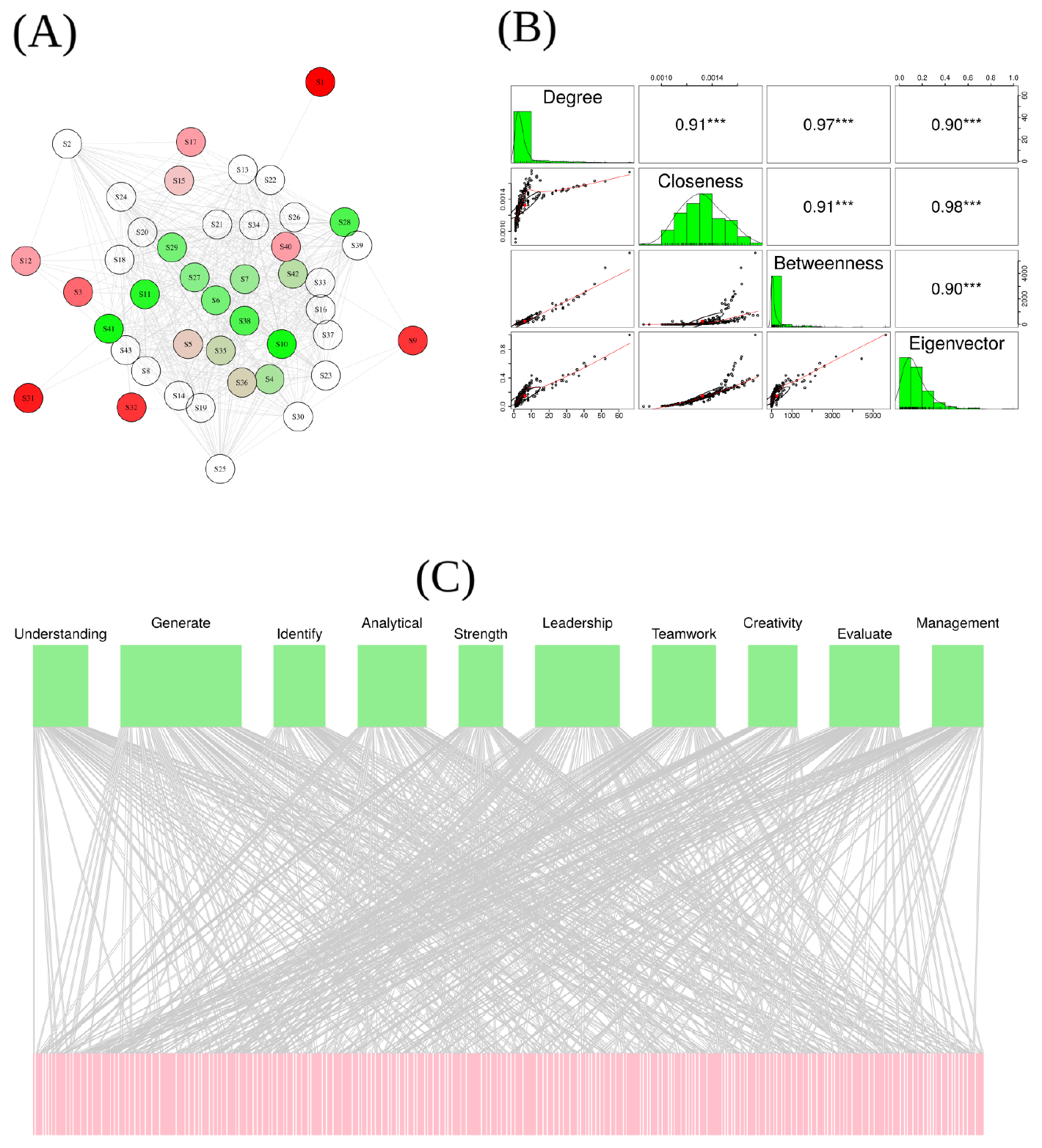}
\end{center}
\caption{\textbf{(A)} Unipartite Projection Network of Emergent Soft Skills from Sampled Graduate Studies.  \textbf{(B)} Spearman correlation matrix plot for the five items of leadership and four items of personality. \textbf{(C)} Bipartite Network of Top-10 soft skills (in green rectangles on the top) emerging from sampled programs (in pink rectangles on the bottom). The symbol *** indicates the correlation is significant at or below 0.001.}
\label{F4}
\end{figure}

We begin our analyses with Figure \ref{F4} and its three panels that help us to describe the statistical network behavior of the resulting 43 terms or n-grams we explicitly searched from the theoretical perspective provided by \cite{Scheerens2020} and \cite{zins2004}. To facilitate the interpretation, we used these 43 terms or n-grams as textual labels of socio-emotional soft skills (see Figure \ref{F4}).

Panel A of Figure \ref{F4} shows these 43 soft skills in three colors. Green nodes in this network proved to be the most central, red nodes were less connected, and white nodes had intermediate connectivity. As expected from the statistical behavior of different centrality measures, panel B of Figure \ref{F4} depicts that the centrality of these skills (i.e., each one showed as a black point in the scatterplot included in the lower triangular matrix) tends to be fairly similar regardless of the mathematical procedure used to estimate soft skills' importance in the bipartite network (i.e., Pearson correlation is greater than or equal to 0.90). In contrast, when the projection of the bipartite network is taken as an input for describing the statistical behavior of soft skills' centrality, we found two highly-correlated centrality metrics (i.e., the correlation between Degree and Eigenvector, and the correlation between Closeness and Betweenness centrality). This finding implies that all centrality measures seem to be good proxies for calculating the importance of soft skills as nodes of a bipartite network but not as nodes of a projected unipartite network.

We used the Eigenvector centrality as our preferred proxy for  soft skills' centrality. Panel C of Figure \ref{F4} shows the bipartite network of top-10 soft skills depicted as green nodes in panel A. These soft skills appear as green rectangles whose width is proportional to their centrality. The straight gray lines connect them with graduate programs, colored in pink rectangles at the bottom. The bipartite network depicted in Panel C of Figure \ref{F4} depicts the general importance of these soft skills. In other words, it shows the top ten soft skills without considering other factors such as the accreditation standard or the program type. In the second part of the Appendix, the reader will find several tables showing a different set of soft skills whose centrality is ranked as a function of these factors.

\begin{figure}[h!]
\begin{center}
\includegraphics[width=1\textwidth]{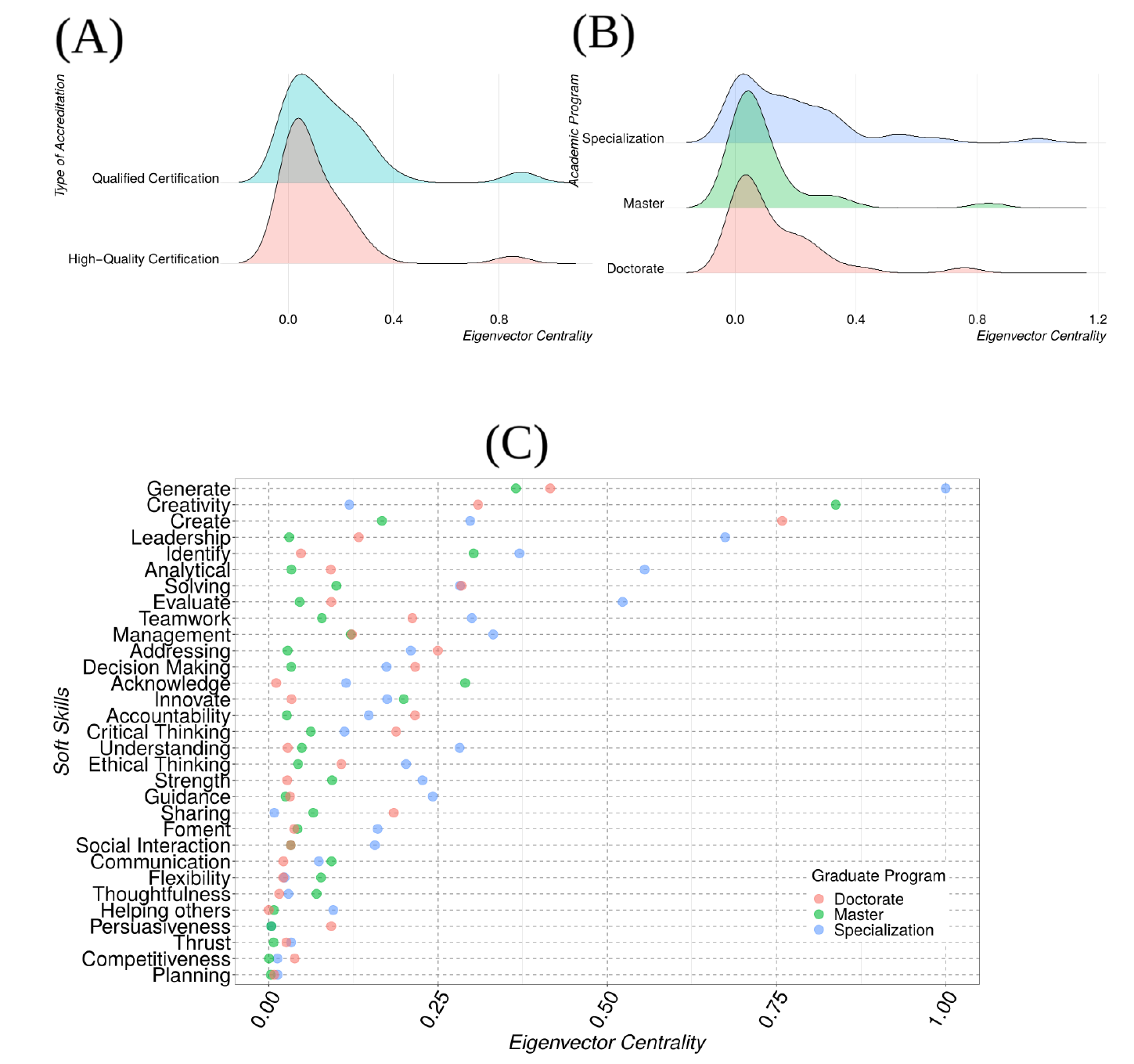}
\end{center}
\caption{\textbf{(A)} Distribution of Soft Skills Centrality by Academic Program.  \textbf{(B)} Distribution of Soft Skills Centrality by Type of Accreditation. \textbf{(C)} List of Transversal Soft Skills' centrality.}
\label{F5}
\end{figure}

To dive deep into these results, we generated five specific term-document matrices (one for each program level plus two for each accreditation type). This computational treatment allowed us to estimate soft skills' centralities better under the assumption that each program responds to the educational needs matching their target students; that is, a specialization program is qualitatively different than a master or a doctorate. We found that although soft skills' centrality varies depending on the accreditation standard (F = 1.49; df = 1; p = 0.226) and the program type (F = 2.558; df = 2; p = 0.0823), their variation differences proved to be statistically non-significant. Panels A and B of Figure \ref{F5} reveal the statistical distribution of soft skills centrality. This result contrasts when it comes to spotting the centrality of each soft skill. Panel C of Figure \ref{F5} shows 31 soft skills explicitly present or commonly shared by specialization, master, and doctorate programs. It is evident that the value of a soft skill centrality is context-dependent in the sense of the sampled documents used as inputs for creating the matrix for the bipartite network representation subjected to analysis.  

\begin{figure}[h!]
\begin{center}
\includegraphics[width=14cm]{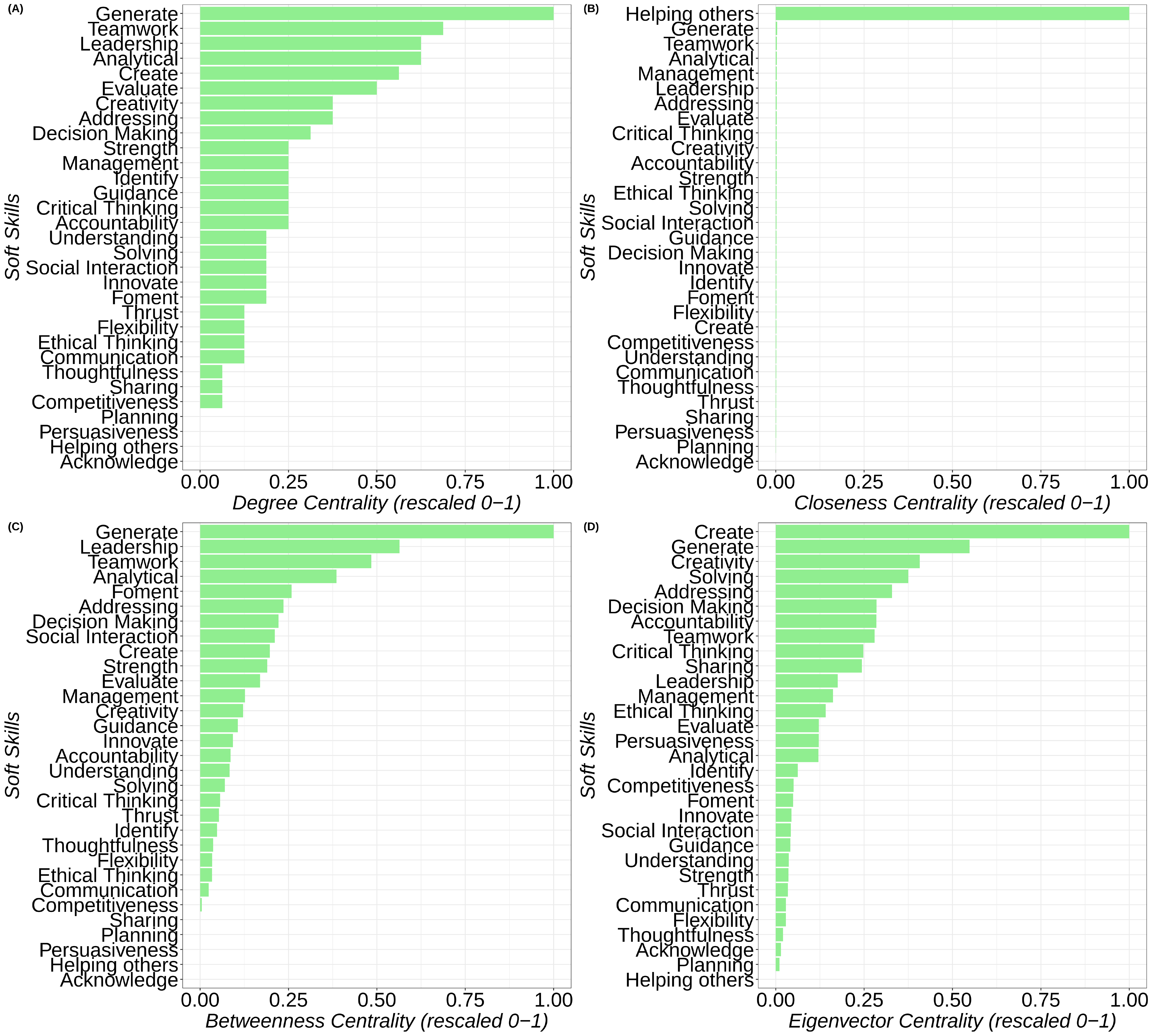}
\end{center}
\caption{Centrality of Soft Skills estimated by \textbf{(A)} degree centrality, \textbf{(B)} Closeness centrality, \textbf{(C)} Betweenness centrality, and \textbf{(D)} Eigenvector centrality. Centrality metrics for all nodes were rescaled to facilitate interpretations.}
\label{F6}
\end{figure}

A final interesting result relates to soft skills' centrality ranking and how some particular soft skills change their position in this ranking depending on the centrality index used. Figure \ref{F6} shows 31 soft skills shared by all academic programs sampled. Soft skills with the highest centrality are those related to creativity (for example, in the creation or generation of ideas or projects), leadership (in the sense of leading others and teamwork), and analytical orientation (for example, evaluating situations and solving problems). In contrast, soft skills with the lowest centrality relate to understanding others, acknowledgment of others, ethical thinking, critical thinking, innovation, accountability, guidance, sharing information, or social interaction.

\section{Discussion}

In the last decade, soft skills have caught the attention of different members of the higher education system \cite{Fletcher2023,Joie2023}. Only recently, however, scholars have attempted to advance our knowledge of soft skills and differentiate them from individual competencies, which in turn, relate to employees' personality traits \cite{Marin2022}. As these soft skills are parts of individual competencies, there seems to be an implicit consensus that ``the interaction between soft skills and training programs appears to be vital to enhancing employee job performance.'' \cite[p. 991]{Marin2022} . In this work, we have extended these ideas with several contributions.

Our first contribution is the conceptual understanding of soft skills from the point of view of graduate studies offerings. As soft skills are thought to improve the scope and impact of employees who know how to use their socio-emotional competencies to facilitate the dynamics at the workplace \cite{Joie2023}, they are considered the missing element in the quality education framework defined by UNESCO's sustainable goals \cite{Jamali2023} and the type of employee required in the twenty-first century \cite{Jang2016,Börner201812630}.

We showed that graduate programs in Colombia tend to emphasize soft skills like creativity (in the sense of creating or generating ideas or projects), leadership (e.g., leading and teamwork), and analytical orientation (e.g., evaluating situations and solving problems). Although these skills seem reasonable regarding what a well-trained professional requires in a concrete technical discipline, we noticed that other skills, such as empathy (i.e., understanding others and acknowledgment of others), ethical thinking, and critical thinking, prove to be less central. Although we regard this result in the context of our exploratory research, it leads us to question if too much emphasis on the most visible soft skills might jeopardize society's long-term wellness. For example, recent evidence shows a dark side of leadership \cite{Benlahcene2022} comprising unethical behaviors that result from ineffective training programs and poor ethics education. As per \cite{Hassan2023511}, unethical leadership relates to behaviors and decisions that are anti-moral, most often illegal, with an outrageous intent to instigate unethical behaviors among employees, producing deleterious impacts on organizations. In the context of higher education, \cite{Crawford20231} poses a timely reflection: ``With the rise of software to make cheating easier, opportunities to outsource dissertations, and a more turbulent sector, it will be the leaders who sustain teams, and build good educational outcomes.'' (p. 1).
Critical thinking with communication is another pair of soft skills that lead us to pose more questions regarding their training in graduate studies. As recent evidence shows that oral and written communication is intrinsically related to critical thinking \cite{hyytinen2023,Campo2023}, we wonder why communication and critical thinking are not as central as creating or generating ideas or projects. How can someone create a project and expect to have collaborators with poor communication skills? Given the exploratory orientation of our work, our findings are far from conclusive. Nonetheless, these results highlight the need to investigate soft skills centrality inside and outside Colombia in more detail. We elaborate upon this idea in the section of limitations and recommendations.

A second contribution of this work is the methodological approach that relies on the combination of bipartite networks with information retrieval and natural processing language techniques and how it offers a promising framework that tackles the problem of quantifying the importance of soft skills in the context of the intended learning outcomes written in the commercial offering of graduate studies. Despite the large body of studies on soft skills, to our knowledge, we are unaware of previous attempts to examine soft skills from this well-known perspective in the emerging field of complex systems \cite{Estrada2011,Correa2020}. This statement, however, should be pondered in the field of higher education research. It is important to recall that bipartite networks are a special case of social networks \cite{Luke2015}, and quite recently, some scholars have leveraged the concept of social network analysis to understand shifts in international student mobility and world university rankings in a time window of twenty years \cite{Glass2023}. From this viewpoint, although the present work provides fresh evidence from Colombia, its methodological approach directly applies to any other nation, opening new horizons to understanding soft skills from graduate studies offerings, which is a sensitive topic for students, university authorities, and policy-makers alike.

The methodological framework provided in this work paves the way for higher education researchers to adopt a data-driven mindset that uses written materials as unstructured data that can be scrutinized following the principles of open science and reproducibility; that is, ``Instead of arguing about whether results hold up, let’s push to provide enough information for others to repeat the experiments'' \cite{Stark2018} (p.613). In higher education research, natural language processing techniques were certainly unknown until recently. For example, \cite{LiCausi2022} used these techniques to understand disciplinary identity in the field of linguistics, and \cite{Mantai2022} used text-mining techniques to identify the expected skills, qualifications, and attributes to do a doctorate. Although the combination of bipartite network analysis with natural language processing techniques is not novel for social and political scientists \cite[see for example,][]{Bail2016}, their combined application for higher education researchers should be explicitly applicable, as illustrated by this work.

A third contribution of this work should be regarded as a corollary of the contributions mentioned above. The content-oriented search and document sampling shown in Figure \ref{F3} open new horizons regarding how to use available data easily accessible through the official websites of HEIs. Higher education researchers, in this sense, are encouraged to explore possible ways to escalate the procedure developed in this work to reach more documents at a continental scale, which might lead to analyzing soft skills centrality from the viewpoint of a cross-national comparison. In this panorama, this work should be regarded as an initial endeavor with a clear research agenda. We took the case of higher education institutions in Colombia, given the academic trajectory of the authors in this country.  As one of the most recent Latin-American members of the OECD, Colombia and its higher education system faces the challenge of increasing their productivity and economic capacities to remain competitive in a globalized market \cite{Zarate2023}. We argue that competitiveness at a national scale relies, among many other factors, on competent individuals in different areas or disciplines for which graduate programs are expected to fill professionals' needs. While the main motivation for professionals who search for a graduate program is to learn new knowledge in a concrete discipline, it is often assumed that this learning might demand some challenges in soft skills and other personal attributes \cite{Suyansah2023,Feraco2023}. Our experience as professors with teaching and research trajectories in Colombia leads us to think that HEIs in this country are particularly demanding for all professionals. Most people enrolled in graduate programs are in the dual role of employees and students, which demands discipline in terms of time management and schedules in a context with few options for sponsoring tuition fees through academic scholarships. If pursuing a master's or a doctorate demands personal accountability to read and assimilate new technical material, but these cognitive tasks are compromised by poor critical thinking, for example, students' attainment and satisfaction might suffer. Thus, we concur with \cite{Mantai2022} that graduate programs ``would do well in communicating how the attributes required for PhD admission will be applied and further developed during the PhD.'' (p. 2283), as empirical evidence shows, among several factors, delays in PhD candidates relate to skills such as communication to prevent supervisors-related and personal-related issues \cite{van2013}.

\subsection{Implications for policy-makers, university authorities, and staff}

The present study offers several implications for policy-makers, university authorities, and staff. With the systematic application of data-driven tools for gathering data monthly, universities can increase their local reputation among different stakeholders by leading the replication of public gatherings similar to those done by the World Economic Forum at the local scale of cities and municipalities. These institutional endeavors can be framed into UNESCO's global sustainability goal of quality education, as they can call for the public participation of students, heads of companies, faculty, and other stakeholders of the higher education system. We concur with \cite{sharvari2019} that technical skills are part of most educational curricula, but soft skills need more emphasis in each syllabus. In other words, educational policies could include explicit guidelines that help university authorities, faculty, and staff include soft skills training in syllabi and graduate studies offerings.

Any attempt to promote mandatory soft skills training in graduate programs will fail if political good intentions are not backed up with insightful, data-driven findings. Thus, refining the criteria employed to identify which soft skills are more critical for each discipline and graduate program is essential. Here, diving deep with more fine-grained research from Colombia and other nations is as vital as desirable. As an initial step, we provided the fundamental elements of a novel conceptual approach to soft skills, hoping to stimulate more evidence for policymakers and university authorities.

It can be helpful to identify the best talent available with a renowned reputation in soft skills training. Professionals with proven interdisciplinary experience can be hired to boost the scope and impact of soft skills training by designing and conducting boot camps that make participants understand why universities should invest time and efforts in redefining all kinds of product and service deliveries of their institutional offerings. University authorities, including presidents or rectors, vice-provosts, faculty deans, and heads of research teams, should be the first participants that attend these boot camps, as these learning experiences will help them understand what young professionals look for and how they enjoy other non-traditional teaching deliveries such as Instagram reels, YouTube videos, smartphone apps, educative video-games, and virtual environments in the metaverse with virtual reality or augmented reality apps. These learning experiences aim to improve the empathy of a generation of professionals that experienced a higher education system that did not rely on these disruptive digital means but understand why such resources can be of great economic value for a new generation of graduate program offerings.

Apart from previous considerations, with this work, we aimed to justify a novel yet promising data analysis procedure that poses some implications for more traditional techniques, such as surveys or questionnaires. Although these data collection techniques will continue to be preferred by several researchers, we argue that they should be regarded as complements or optional means to collect relevant data on soft skills. Thus, we foresee a rise in natural language processing techniques as data collection procedures for soft skills research. Although data collection was certainly mentioned as a related factor to soft skills research in recent reviews like the one provided by \cite{Marin2022}, the absence of bipartite networks and natural language processing in such endeavors is remarkable. A practical implication from this perspective is to take advantage of the reproducible research writing style provided in this work to apply this procedure in other contexts and countries.

\subsection{Limitations and recommendations for future research}

This work is not free of limitations. First, the exploratory character of our work makes our findings should be regarded as preliminary as seminal for further research. Future endeavors can follow at least two of the following complementary routes. On the one hand, examining soft skills centrality for a specific graduate program offered by different universities might be highly relevant. For example, by collecting a sample of graduate programs in mechanical engineering, one can easily define the set of specific soft skills that works as the core for such a profession regardless of the institution where the students decide to enroll. On the other hand, it is essential to conduct a similar effort on a sample of graduate programs that share common disciplinary knowledge with different emphases, such as psychology and its sub-disciplines like consumer psychology, clinical psychology, health psychology, social psychology, etc. The two routes mentioned here can help understand how professional associations might serve as external resources for soft skills training outside traditional classroom formats and becoming strategic partners of HEIs.

Second, the focus on Colombia surely reveals some of the features of the higher education system of this country, and these characteristics are not necessarily similar to those of other nations, even with similar cultural and historical roots. Further investigations from other countries are highly recommended and welcome, as they can provide evidence of soft skills centrality as a benchmark.

\section{Conclusion}

An increasing challenge for HEIs relates to soft skills training in their graduate programs. The idea to improve graduate studies with such training is aligned with UNESCO's sustainable goal of providing quality education so that professionals of any discipline can join a job market with a set of socio-emotional competencies that allow them to boost their performance at the workplace. Our exploratory work has shown a fruitful conceptual and methodological approach that paves the way for others to dive deep into training soft skills that are highly central for disciplinary graduate programs. As the development of soft skills should not be restrictive or exclusively oriented toward graduate students, universities can help to extend the training of these skills from the viewpoint of the so-called long-life learning and provide alternatives for undergraduates, teenagers, and toddlers, as well as senior executives and consultants, and retired personnel.

\newpage
\section*{Appendix}
This appendix has two parts. This first part illustrates the computational details we developed to proceed with the analysis. The following code provides a step-by-step procedure that is helpful for individuals who are not familiarized with R as a programming language and statistical analysis tool. The first part serves as guidance for identifying and estimating the soft skills' centrality. The second part provides detailed lists of top ten soft skills by program type and accreditation standard. These two parts aim to complement the section on materials and methods and facilitate an understanding of how we reached the findings reported in the section on results.

\section*{Part 1: Computational details}
This code was run under the following computer and RStudio session info.\\

\noindent
\scriptsize{
\begin{verbatim}
# Step 1: Opening the sample of texts
# this local folder is a clone of the GitHub Repo
setwd("/home/jc/Documents/Paper Soft Skills Sampled Programs")
listado <- data.frame(dir())
library(readtext)
library(tm)
DirSource()
# Get the data directory from readtext
DATA_DIR <- system.file("extdata/", package = "readtext")
textos <- readtext(listado$dir..)
textos$doc_id <- gsub("[^0-9-]", "", textos$doc_id)


# Step 2: Creating a corpus from texts
library(quanteda)
Textos <- corpus(textos)


# Step 3: Tagging the texts according to
# their program type and accreditation
source("~/Documents/GitHub/SoftSkillsUniversityPrograms/SampleAnalysis.R")
docvars(Textos, "Programa") <- Muestra$NOMBRE_DEL_PROGRAMA
docvars(Textos, "Program.Level") <- Muestra$Programa
docvars(Textos, "Institution") <- Muestra$NOMBRE_INSTITUCION
docvars(Textos, "Accreditation") <- Muestra$Accreditation
summary(Textos)
aja <- data.frame(summary(Textos, n = length(Textos)))
SPEC <- corpus_subset(Textos, Program.Level == "Specialization")
MS <- corpus_subset(Textos, Program.Level == "Masters")
PhD <- corpus_subset(Textos, Program.Level == "Doctorate")
QC <- corpus_subset(Textos, Accreditation == "Qualified Certification")
HQC <- corpus_subset(Textos, Accreditation == "High-Quality Certification")
phd <- data.frame(summary(PhD, n = length(PhD)))

# Step 4. Soft Skills theoretically driven identification
# Keywords-in-context Search
pc <- data.frame(kwic(Textos, pattern = phrase("pensamiento critico")))
sp <- data.frame(kwic(Textos, pattern = phrase("solucionar problemas")))
comunicar <- data.frame(kwic(Textos, pattern = "comunicar"))
creatividad <- data.frame(kwic(Textos, pattern = "creatividad"))
paciencia <- data.frame(kwic(Textos, pattern = "paciencia"))
crear <- data.frame(kwic(Textos, pattern = "crear"))
liderar <- data.frame(kwic(Textos, pattern = "liderar"))
resolver <- data.frame(kwic(Textos, pattern = "resolver"))
comprometer <- data.frame(kwic(Textos, pattern = "comprometer"))
comprometerse <- data.frame(kwic(Textos, pattern = "comprometerse"))
gestionar <- data.frame(kwic(Textos, pattern = "gestionar"))
reflexionar <- data.frame(kwic(Textos, pattern = "reflexionar"))
controlar <- data.frame(kwic(Textos, pattern = "controlar"))
etico <- data.frame(kwic(Textos, pattern = "etico"))
tolerar <- data.frame(kwic(Textos, pattern = "tolerar"))
argumentar <- data.frame(kwic(Textos, pattern = "argumentar"))
conflicto <- data.frame(kwic(Textos, pattern = "conflictos"))
negociar <- data.frame(kwic(Textos, pattern = "negociar"))
comprender <- data.frame(kwic(Textos, pattern = "comprender"))
equipo <- data.frame(kwic(Textos, pattern = "equipos"))
planificar <- data.frame(kwic(Textos, pattern = "planificar"))
generar <- data.frame(kwic(Textos, pattern = "generar"))
empatia <- data.frame(kwic(Textos, pattern = "empatia"))
compartir <- data.frame(kwic(Textos, pattern = "compartir"))
analizar <- data.frame(kwic(Textos, pattern = "analizar"))
reconocer <- data.frame(kwic(Textos, pattern = "reconocer"))
orientar <- data.frame(kwic(Textos, pattern = "orientar"))
respetar <- data.frame(kwic(Textos, pattern = "respetar"))
motivar <- data.frame(kwic(Textos, pattern = "motivar"))
cooperar <- data.frame(kwic(Textos, pattern = "cooperar"))
fortalecer <- data.frame(kwic(Textos, pattern = "fortalecer"))
impulsar <- data.frame(kwic(Textos, pattern = "impulsar"))
acercar <- data.frame(kwic(Textos, pattern = "acercar"))
ayudar <- data.frame(kwic(Textos, pattern = "ayudar"))
cambiar <- data.frame(kwic(Textos, pattern = "cambiar"))
apreciar <- data.frame(kwic(Textos, pattern = "apreciar"))
dirigir <- data.frame(kwic(Textos, pattern = "dirigir"))
fomentar <- data.frame(kwic(Textos, pattern = "fomentar"))
interactuar <- data.frame(kwic(Textos, pattern = "interactuar"))
identificar <- data.frame(kwic(Textos, pattern = "identificar"))
competir <- data.frame(kwic(Textos, pattern = "competir"))
manifestar <- data.frame(kwic(Textos, pattern = "manifestar"))
responsable <- data.frame(kwic(Textos, pattern = "responsable"))
evaluar <- data.frame(kwic(Textos, pattern = "evaluar"))
innovar <- data.frame(kwic(Textos, pattern = "innovar"))
decidir <- data.frame(kwic(Textos, pattern = "decidir"))
td <- data.frame(kwic(Textos, pattern = phrase("tomar decisiones")))
flex <- data.frame(kwic(Textos, pattern = "flexibilidad"))
persu <- data.frame(kwic(Textos, pattern = "persua*"))
conven <- data.frame(kwic(Textos, pattern = "convencer"))

rm(institution, LevelsOfficials,
   LevelsPrivate, listado, Muestra,
   Officials, Private, Sector, textos,
   Textos, DATA_DIR)
TODAS <- rbind(persu, conven, flex, td, decidir, sp,
               pc, creatividad, paciencia, crear,
               innovar, acercar, analizar, apreciar,
               argumentar, ayudar, cambiar, compartir,
               competir, comprender, comprometer,
               comprometerse, comunicar, conflicto,
               controlar, cooperar, dirigir, empatia,
               equipo, etico, evaluar, fomentar, fortalecer,
               generar, gestionar, identificar, impulsar,
               interactuar, liderar, manifestar, motivar,
               negociar, orientar, planificar, reconocer,
               reflexionar, resolver, respetar,
               responsable, tolerar)

colnames(aja)[1] <- "docname"
library(dplyr)
TODAS2 <- TODAS %>%
  select(-from, -to, -pre, -post, -pattern) %>%
  left_join(aja, by = "docname")
Spec <- TODAS2 %>% filter(., Program.Level == "Specialization")
MS <- TODAS2 %>% filter(., Program.Level == "Masters")
PhD <- TODAS2 %>% filter(., Program.Level == "Doctorate")

# Step 5. Plotting results
load("~/Documents/GitHub/SoftSkillsUniversityPrograms/DataForFigure4.RData")
rm(list=setdiff(ls(), "TODAS"))
TODAS[TODAS=="acercar"] <- "S1"
TODAS[TODAS=="analizar"] <- "S2"
TODAS[TODAS=="argumentar"] <- "S3"
TODAS[TODAS=="ayudar"] <- "S4"
TODAS[TODAS=="cambiar"] <- "S5"
TODAS[TODAS=="compartir"] <- "S6"
TODAS[TODAS=="competir"] <- "S7"
TODAS[TODAS=="comprender"] <- "S8"
TODAS[TODAS=="comprometerse"] <- "S9"
TODAS[TODAS=="comunicar"] <- "S10"
TODAS[TODAS=="conflictos"] <- "S11"
TODAS[TODAS=="controlar"] <- "S12"
TODAS[TODAS=="crear"] <- "S13"
TODAS[TODAS=="creatividad"] <- "S14"
TODAS[TODAS=="decidir"] <- "S15"
TODAS[TODAS=="dirigir"] <- "S16"
TODAS[TODAS=="empatı́a"] <- "S17"
TODAS[TODAS=="equipos"] <- "S18"
TODAS[TODAS=="ético"] <- "S19"
TODAS[TODAS=="evaluar"] <- "S20"
TODAS[TODAS=="flexibilidad"] <- "S21"
TODAS[TODAS=="fomentar"] <- "S22"
TODAS[TODAS=="fortalecer"] <- "S23"
TODAS[TODAS=="generar"] <- "S24"
TODAS[TODAS=="gestionar"] <- "S25"
TODAS[TODAS=="identificar"] <- "S26"
TODAS[TODAS=="impulsar"] <- "S27"
TODAS[TODAS=="innovar"] <- "S28"
TODAS[TODAS=="interactuar"] <- "S29"
TODAS[TODAS=="liderar"] <- "S30"
TODAS[TODAS=="manifestar"] <- "S31"
TODAS[TODAS=="motivar"] <- "S32"
TODAS[TODAS=="orientar"] <- "S33"
TODAS[TODAS=="pensamiento crı́tico"] <- "S34"
TODAS[TODAS=="persuasión"] <- "S35"
TODAS[TODAS=="planificar"] <- "S36"
TODAS[TODAS=="reconocer"] <- "S37"
TODAS[TODAS=="reflexionar"] <- "S38"
TODAS[TODAS=="resolver"] <- "S39"
TODAS[TODAS=="respetar"] <- "S40"
TODAS[TODAS=="responsable"] <- "S41"
TODAS[TODAS=="solucionar problemas"] <- "S42"
TODAS[TODAS=="tomar decisiones"] <- "S43"

# Figure 4 Panel A
Network <- TODAS[,c(1,5)]
table(Network$keyword)
Network <- Network[!duplicated(Network[c(1,2)]),]
library(igraph)
bn2 <- graph.data.frame(Network,directed=FALSE)
bipartite.mapping(bn2)
V(bn2)$type <- bipartite_mapping(bn2)$type
V(bn2)$color <- ifelse(V(bn2)$type, "red", "green")
V(bn2)$shape <- ifelse(V(bn2)$type, "circle", "square")
V(bn2)$label.cex <- ifelse(V(bn2)$type, 0.5, 1)
V(bn2)$size <- sqrt(igraph::degree(bn2))
E(bn2)$color <- "lightgrey"
bn2.pr <- bipartite.projection(bn2)
Terms <- bn2.pr$proj2
centrality_scores <- degree(Terms)
CS <- centrality_scores
# Normalize the centrality scores to a range between 0 and 1, 
# as follows:
# centrality_scores - min(centrality_scores) (in the numerator)
# (max(centrality_scores) - min(centrality_scores) (in the denominator)


normalized_scores <- (CS - min(CS)) / (max(CS) - min(CS))

# Create a color palette with different colors
color_palette <- colorRampPalette(c("red", "pink", "lightgreen", "green"))
(length(unique(normalized_scores)))

# Assign colors to nodes based on their normalized centrality scores
node_colors <- color_palette[rank(normalized_scores)]

# Plot the network with node colors based on centrality
plot(Terms, vertex.label.color = "black", 
     vertex.label.cex = 0.8, 
     vertex.color = node_colors, 
     vertex.size = 15, edge.width = 0.5, 
     edge.color = "lightgray", 
     layout = layout_components, main = "")

# Figure 4 Panel B
load("~/Documents/GitHub/SoftSkillsUniversityPrograms/DataForFigure4.RData")
rm(list=setdiff(ls(), "SkillsProgramsCentrality"))

library(psych)
pairs.panels(SkillsProgramsCentrality, 
             method = "spearman", 
             hist.col = "green",
             density = TRUE,  
             ellipses = TRUE,
             pch = 21,
             cex = 2.5,
             cex.axis = 1.8,
             cex.labels = 4.5,
             lwd = 2,
             rug = TRUE,
             stars = TRUE
)

# Figure 4 Panel C
load("/home/jc/Documents/GitHub/SoftSkillsUniversityPrograms")
rm(list=setdiff(ls(), "DTM3"))
# The DTM3 object is a matrix with 10 columns (with the soft skills 
# that proved to be more central and all programs as rows. In this
# matrix several programs don't have a connection with any of these
# central skills. Thus, we will discard these programs to decpict
# a bipartite Network for illustrative purposes.)

DTM4 <- apply(DTM3, 1, function(row) any(row != 0))
BN <- DTM3[DTM4, ]

library(bipartite)
plotweb(BN, method = "normal", 
        col.high = "lightgreen", 
        bor.col.high = "lightgreen",
        col.low = "pink", 
        bor.col.low = "pink",
        col.interaction = "grey90",
        bor.col.interaction = "grey90",
        low.lablength = 0,
        labsize = 2)


# Figure 5A & 5B
library(ggplot2)
library(ggridges)
p1 <- ggplot(Centralities, aes(x = Eigen.vector, y = Level, fill = Level)) +
  geom_density_ridges(alpha = 0.3) +
  theme_ridges() + 
  theme(legend.position = "none") + 
  xlab("Eigenvector Centrality") + 
  ylab("Academic Program") + 
  theme(axis.text.x=element_text(size=35)) +
  theme(axis.text.y=element_text(size=35)) +
  theme(axis.title.x=element_text(face="italic", colour="black", size=35)) +
  theme(axis.title.y=element_text(face="italic", colour="black", size=35))


p2 <- ggplot(Centralities2, aes(x = Eigen.vector, y = Accreditation, 
  fill = Accreditation)) +
  geom_density_ridges(alpha = 0.3) +
  theme_ridges() + 
  theme(legend.position = "none") + 
  xlab("Eigenvector Centrality") + 
  ylab("Type of Accreditation") + 
  theme(axis.text.x=element_text(size=35)) +
  theme(axis.text.y=element_text(size=35)) +
  theme(axis.title.x=element_text(face="italic", colour="black", size=35)) +
  theme(axis.title.y=element_text(face="italic", colour="black", size=35))


library(ggpubr)
ggarrange(p1, p2, labels = c("(A)", "(B)"), ncol = 1, nrow = 2)


# Figure 5C
# Here we found that 31 soft skills are present all across sampled programs
table(Resumen$Freq)
library(dplyr)
SoftSkillsCentrality <- Centralities %>% 
  filter(., grepl('analizar|ayudar|
                  compartir|competir|
                  comprender|comunicar|
                  crear|creatividad|dirigir|
                  equipos|ético|evaluar|
                  flexibilidad|fomentar|
                  fortalecer|generar|gestionar|
                  identificar|impulsar|innovar|
                  interactuar|liderar|orientar|
                  pensamiento crítico|persuasión|
                  planificar|reconocer|reflexionar|
                  resolver|responsable|tomar decisiones', SS))

boxplot(SoftSkillsCentrality$Closeness, xlab = "Closeness") 
boxplot(SoftSkillsCentrality$Betweennes, xlab = "Betweennes") 
boxplot(SoftSkillsCentrality$Degree,  xlab = "Degree")
boxplot(SoftSkillsCentrality$Eigen.vector, xlab = "Eigenvector")


boxplot(SoftSkillsCentrality[c(2,4)])


library(ggplot2)
ggplot(SoftSkillsCentrality, aes(x=reorder(SS, Closeness), y=Closeness)) +
  scale_fill_discrete(name="Academic Program") + 
  geom_point(size=5, aes(colour=Level), alpha=0.6) +
  # Use a larger dot
  theme_bw() +
  theme(axis.text.x = element_text(angle=60, hjust=1),
        panel.grid.major.y = element_line(colour="grey60", linetype="dashed"),
        panel.grid.minor.y = element_blank(),
        panel.grid.major.x = element_line(colour="grey60", linetype="dashed"),) +
  coord_flip() + theme(legend.position="top") +
  theme(axis.text.x=element_text(size=15, colour="black")) +
  theme(axis.text.y=element_text(size=15, colour="black")) +
  theme(axis.title.x=element_text(face="italic", colour="black", size=20)) +
  theme(axis.title.y=element_text(face="italic", colour="black", size=20)) +
  xlab("Soft Skills") +
  ylab("Closeness Centrality") +
  theme(legend.position=c(0.95,0.1), legend.justification=c(0.95,0.1)) 


SoftSkillsCentrality$SS[SoftSkillsCentrality$SS == 'generar'] <- 'Generate'
SoftSkillsCentrality$SS[SoftSkillsCentrality$SS == 'creatividad'] <- 'Creativity'
SoftSkillsCentrality$SS[SoftSkillsCentrality$SS == 'crear'] <- 'Create'
SoftSkillsCentrality$SS[SoftSkillsCentrality$SS == 'liderar'] <- 'Leadership'
SoftSkillsCentrality$SS[SoftSkillsCentrality$SS == 'identificar'] <- 'Identify'
SoftSkillsCentrality$SS[SoftSkillsCentrality$SS == 'analizar'] <- 'Analytical'
SoftSkillsCentrality$SS[SoftSkillsCentrality$SS == 'resolver'] <- 'Solving'
SoftSkillsCentrality$SS[SoftSkillsCentrality$SS == 'evaluar'] <- 'Evaluate'
SoftSkillsCentrality$SS[SoftSkillsCentrality$SS == 'equipos'] <- 'Teamwork'
SoftSkillsCentrality$SS[SoftSkillsCentrality$SS == 'gestionar'] <- 'Management'
SoftSkillsCentrality$SS[SoftSkillsCentrality$SS == 'dirigir'] <- 'Addressing'
SoftSkillsCentrality$SS[SoftSkillsCentrality$SS == 'tomar decisiones'] <- 'Decision Making'
SoftSkillsCentrality$SS[SoftSkillsCentrality$SS == 'reconocer'] <- 'Acknowledge'
SoftSkillsCentrality$SS[SoftSkillsCentrality$SS == 'innovar'] <- 'Innovate'
SoftSkillsCentrality$SS[SoftSkillsCentrality$SS == 'responsable'] <- 'Accountability'
SoftSkillsCentrality$SS[SoftSkillsCentrality$SS == 'pensamiento crítico'] <- 'Critical Thinking'
SoftSkillsCentrality$SS[SoftSkillsCentrality$SS == 'comprender'] <- 'Understanding'
SoftSkillsCentrality$SS[SoftSkillsCentrality$SS == 'ético'] <- 'Ethical Thinking'
SoftSkillsCentrality$SS[SoftSkillsCentrality$SS == 'fortalecer'] <- 'Strength'
SoftSkillsCentrality$SS[SoftSkillsCentrality$SS == 'orientar'] <- 'Guidance'
SoftSkillsCentrality$SS[SoftSkillsCentrality$SS == 'compartir'] <- 'Sharing'
SoftSkillsCentrality$SS[SoftSkillsCentrality$SS == 'fomentar'] <- 'Foment'
SoftSkillsCentrality$SS[SoftSkillsCentrality$SS == 'interactuar'] <- 'Social Interaction'
SoftSkillsCentrality$SS[SoftSkillsCentrality$SS == 'comunicar'] <- 'Communication'
SoftSkillsCentrality$SS[SoftSkillsCentrality$SS == 'flexibilidad'] <- 'Flexibility'
SoftSkillsCentrality$SS[SoftSkillsCentrality$SS == 'reflexionar'] <- 'Thoughtfulness'
SoftSkillsCentrality$SS[SoftSkillsCentrality$SS == 'ayudar'] <- 'Helping others'
SoftSkillsCentrality$SS[SoftSkillsCentrality$SS == 'persuasión'] <- 'Persuasiveness'
SoftSkillsCentrality$SS[SoftSkillsCentrality$SS == 'impulsar'] <- 'Thrust'
SoftSkillsCentrality$SS[SoftSkillsCentrality$SS == 'competir'] <- 'Competitiveness'
SoftSkillsCentrality$SS[SoftSkillsCentrality$SS == 'planificar'] <- 'Planning'

library(ggplot2)
p <- ggplot(SoftSkillsCentrality, aes(x=reorder(SS, Eigen.vector), y=Eigen.vector)) +
  scale_fill_discrete(name="Academic Program") + 
  geom_point(size=5, aes(colour=Level), alpha=0.6) +
  # Use a larger dot
  theme_bw() + 
  theme(axis.text.x = element_text(angle=60, hjust=1),
        panel.grid.major.y = element_line(colour="grey60", linetype="dashed"),
        panel.grid.minor.y = element_blank(),
        panel.grid.major.x = element_line(colour="grey60", linetype="dashed"),) +
  coord_flip() + theme(legend.position="top") +
  theme(axis.text.x=element_text(size=25, colour="black")) +
  theme(axis.text.y=element_text(size=25, colour="black")) +
  theme(axis.title.x=element_text(face="italic", colour="black", size=25)) +
  theme(axis.title.y=element_text(face="italic", colour="black", size=25)) +
  xlab("Soft Skills") +
  ylab("Eigenvector Centrality") +
  theme(legend.title=element_text(size=20), 
        legend.text = element_text(size = 20), 
        legend.position=c(0.95,0.1), 
        legend.justification=c(0.95,0.1)) 

p + labs(color = "Program Type")

# Figure 6
library(dplyr)
SoftSkillsCentrality <- Centralities %>% 
  filter(., grepl('analizar|ayudar|compartir|
                  competir|comprender|comunicar|
                  crear|creatividad|dirigir|equipos|
                  etico|evaluar|flexibilidad|fomentar|
                  fortalecer|generar|gestionar|
                  identificar|impulsar|innovar|interactuar|
                  liderar|orientar|pensamiento crítico|
                  persuasión|planificar|reconocer|reflexionar|
                  resolver|responsable|tomar decisiones', SS))

SoftSkillsCentrality$SS[SoftSkillsCentrality$SS == 'generar'] <- 'Generate'
SoftSkillsCentrality$SS[SoftSkillsCentrality$SS == 'creatividad'] <- 'Creativity'
SoftSkillsCentrality$SS[SoftSkillsCentrality$SS == 'crear'] <- 'Create'
SoftSkillsCentrality$SS[SoftSkillsCentrality$SS == 'liderar'] <- 'Leadership'
SoftSkillsCentrality$SS[SoftSkillsCentrality$SS == 'identificar'] <- 'Identify'
SoftSkillsCentrality$SS[SoftSkillsCentrality$SS == 'analizar'] <- 'Analytical'
SoftSkillsCentrality$SS[SoftSkillsCentrality$SS == 'resolver'] <- 'Solving'
SoftSkillsCentrality$SS[SoftSkillsCentrality$SS == 'evaluar'] <- 'Evaluate'
SoftSkillsCentrality$SS[SoftSkillsCentrality$SS == 'equipos'] <- 'Teamwork'
SoftSkillsCentrality$SS[SoftSkillsCentrality$SS == 'gestionar'] <- 'Management'
SoftSkillsCentrality$SS[SoftSkillsCentrality$SS == 'dirigir'] <- 'Addressing'
SoftSkillsCentrality$SS[SoftSkillsCentrality$SS == 'tomar decisiones'] <- 'Decision Making'
SoftSkillsCentrality$SS[SoftSkillsCentrality$SS == 'reconocer'] <- 'Acknowledge'
SoftSkillsCentrality$SS[SoftSkillsCentrality$SS == 'innovar'] <- 'Innovate'
SoftSkillsCentrality$SS[SoftSkillsCentrality$SS == 'responsable'] <- 'Accountability'
SoftSkillsCentrality$SS[SoftSkillsCentrality$SS == 'pensamiento crítico'] <- 'Critical Thinking'
SoftSkillsCentrality$SS[SoftSkillsCentrality$SS == 'comprender'] <- 'Understanding'
SoftSkillsCentrality$SS[SoftSkillsCentrality$SS == 'etico'] <- 'Ethical Thinking'
SoftSkillsCentrality$SS[SoftSkillsCentrality$SS == 'fortalecer'] <- 'Strength'
SoftSkillsCentrality$SS[SoftSkillsCentrality$SS == 'orientar'] <- 'Guidance'
SoftSkillsCentrality$SS[SoftSkillsCentrality$SS == 'compartir'] <- 'Sharing'
SoftSkillsCentrality$SS[SoftSkillsCentrality$SS == 'fomentar'] <- 'Foment'
SoftSkillsCentrality$SS[SoftSkillsCentrality$SS == 'interactuar'] <- 'Social Interaction'
SoftSkillsCentrality$SS[SoftSkillsCentrality$SS == 'comunicar'] <- 'Communication'
SoftSkillsCentrality$SS[SoftSkillsCentrality$SS == 'flexibilidad'] <- 'Flexibility'
SoftSkillsCentrality$SS[SoftSkillsCentrality$SS == 'reflexionar'] <- 'Thoughtfulness'
SoftSkillsCentrality$SS[SoftSkillsCentrality$SS == 'ayudar'] <- 'Helping others'
SoftSkillsCentrality$SS[SoftSkillsCentrality$SS == 'persuasión'] <- 'Persuasiveness'
SoftSkillsCentrality$SS[SoftSkillsCentrality$SS == 'impulsar'] <- 'Thrust'
SoftSkillsCentrality$SS[SoftSkillsCentrality$SS == 'competir'] <- 'Competitiveness'
SoftSkillsCentrality$SS[SoftSkillsCentrality$SS == 'planificar'] <- 'Planning'


dat <- SoftSkillsCentrality[1:5]
options(scipen = 999)
dat <- SoftSkillsCentrality %>% filter(., Level == "Doctorate")

library(datawizard)
dat <- dat %>% 
mutate(., degree.rescaled = ifelse(Degree == 0, 0.00, rescale(dat$Degree, to = c(0,1))))
dat <- dat %>% 
mutate(., closeness.rescaled = ifelse(Closeness == 0, 0.00, rescale(dat$Closeness, to = c(0,1))))
dat <- dat %>% 
mutate(., betweennes.rescaled = ifelse(Betweennes == 0, 0.00, rescale(dat$Betweennes, to = c(0,1))))
dat <- dat %>% 
mutate(., eigenvector.rescaled = ifelse(Eigen.vector == 0, 0.00, rescale(dat$Eigen.vector, to = c(0,1))))

summary(dat$eigenvector.rescaled)
summary(dat$degree.rescaled)
summary(dat$betweennes.rescaled)
summary(dat$closeness.rescaled)
colnames(dat)


p1 <- ggplot(dat, aes(x = reorder(SS, degree.rescaled), y = degree.rescaled)) +
  geom_bar(stat = "identity", fill="lightgreen") + theme_bw() + 
  theme(axis.text.x=element_text(size=25, colour="black")) +
  theme(axis.text.y=element_text(size=25, colour="black")) +
  theme(axis.title.x=element_text(face="italic", colour="black", size=25)) +
  theme(axis.title.y=element_text(face="italic", colour="black", size=25)) +
  coord_flip() + xlab("Soft Skills") + ylab("Degree Centrality (rescaled 0-1)")

p2 <- ggplot(dat, aes(x = reorder(SS, closeness.rescaled), y = closeness.rescaled)) +
  geom_bar(stat = "identity", fill="lightgreen") + theme_bw() + 
  theme(axis.text.x=element_text(size=25, colour="black")) +
  theme(axis.text.y=element_text(size=25, colour="black")) +
  theme(axis.title.x=element_text(face="italic", colour="black", size=25)) +
  theme(axis.title.y=element_text(face="italic", colour="black", size=25)) +
  coord_flip() + xlab("Soft Skills") + ylab("Closeness Centrality (rescaled 0-1)")

p3 <- ggplot(dat, aes(x = reorder(SS, betweennes.rescaled), y = betweennes.rescaled)) +
  geom_bar(stat = "identity", fill="lightgreen") + theme_bw() +
  theme(axis.text.x=element_text(size=25, colour="black")) +
  theme(axis.text.y=element_text(size=25, colour="black")) +
  theme(axis.title.x=element_text(face="italic", colour="black", size=25)) +
  theme(axis.title.y=element_text(face="italic", colour="black", size=25)) +
  coord_flip() + xlab("Soft Skills") + ylab("Betweenness Centrality (rescaled 0-1)")

p4 <- ggplot(dat, aes(x = reorder(SS, eigenvector.rescaled), y = eigenvector.rescaled)) +
  geom_bar(stat = "identity", fill="lightgreen") + theme_bw() + 
  theme(axis.text.x=element_text(size=25, colour="black")) +
  theme(axis.text.y=element_text(size=25, colour="black")) +
  theme(axis.title.x=element_text(face="italic", colour="black", size=25)) +
  theme(axis.title.y=element_text(face="italic", colour="black", size=25)) +
  coord_flip() + xlab("Soft Skills") + ylab("Eigenvector Centrality (rescaled 0-1)")



library(ggpubr)
figure <- ggarrange(p1, p2, p3, p4, 
                    labels = c("(A)", "(B)", "(C)", "(D)"),
                    ncol = 2, nrow = 2)

figure



# Generating Tables for Appendix

load("~/Documents/GitHub/SoftSkillsUniversityPrograms/ResultsbyProgram.RData")
load("~/Documents/GitHub/SoftSkillsUniversityPrograms/ResultsbyAccreditation.RData")
library(tidyverse)
Top10Specialization <- SoftSkillsCentrality %>% 
  filter(,SoftSkillsCentrality$Level == "Specialization") %>% 
  select(, SS, Eigen.vector) %>% 
  arrange(, desc(Eigen.vector))

Spec <- head(Top10Specialization, 10)

library(xtable)
spec <- xtable(Spec)
print(spec, include.rownames = TRUE, floating = FALSE, tabular.environment = "longtable")

Top10Master <- SoftSkillsCentrality %>% 
  filter(,SoftSkillsCentrality$Level == "Master") %>% 
  select(, SS, Eigen.vector) %>% 
  arrange(, desc(Eigen.vector))

MS <- head(Top10Master, 10)

master <- xtable(MS)
print(master, include.rownames = TRUE, floating = FALSE, tabular.environment = "longtable")


Top10PhD <- SoftSkillsCentrality %>% 
  filter(,SoftSkillsCentrality$Level == "Doctorate") %>% 
  select(, SS, Eigen.vector) %>% 
  arrange(, desc(Eigen.vector))

PhD <- head(Top10PhD, 10)

phd <- xtable(PhD)
print(phd, include.rownames = TRUE, floating = FALSE, tabular.environment = "longtable")

Top10HQS <- Centralities2 %>% 
  filter(,Accreditation == "High-Quality Certification") %>% 
  select(, SS, Eigen.vector) %>% 
  arrange(, desc(Eigen.vector))

HQS <- head(Top10HQS, 10)

hqs <- xtable(HQS)
print(hqs, include.rownames = TRUE, floating = FALSE, tabular.environment = "longtable")


Top10QC <- Centralities2 %>% 
  filter(,Accreditation == "Qualified Certification") %>% 
  select(, SS, Eigen.vector) %>% 
  arrange(, desc(Eigen.vector))

QC <- head(Top10QC, 10)

qc <- xtable(QC)
print(qc, include.rownames = TRUE, floating = FALSE, tabular.environment = "longtable")
\end{verbatim}

\section*{Part 2: Top-ten soft skills list by program type and accreditation standard}
\begin{longtable}{llr}
  \caption{Top Ten Soft Skills for Specialization Programs}\\
  \hline
Rank & Soft Skill & Eigen.vector \\ 
  \hline
1 & Generate & 1.00 \\ 
2 & Leadership & 0.67 \\ 
3 & Analytical & 0.56 \\ 
4 & Evaluate & 0.52 \\ 
5 & Identify & 0.37 \\ 
6 & Management & 0.33 \\ 
7 & Teamwork & 0.30 \\ 
8 & Create & 0.30 \\ 
9 & Solving & 0.28 \\ 
10 & Understanding & 0.28 \\ 
   \hline
\hline
\end{longtable}


\begin{longtable}{llr}
\caption{Top Ten Soft Skills for Master's Programs}\\
\hline\\
Rank & Soft Skill & Eigen.vector \\ 
  \hline
1 & Creativity & 0.84 \\ 
2 & Generate & 0.37 \\ 
3 & Identify & 0.30 \\ 
4  & Acknowledge & 0.29 \\ 
5  & Innovate & 0.20 \\ 
6  & Create & 0.17 \\ 
7  & Management & 0.12 \\ 
8  & Solving & 0.10 \\ 
9  & Strength & 0.09 \\ 
10 & Communication & 0.09 \\ 
   \hline
\hline
\end{longtable}

\begin{longtable}{llr}
\caption{Top Ten Soft Skills for Doctorate Programs}\\
\hline
Rank & Soft Skill & Eigen.vector \\ 
  \hline
1 & Create & 0.76 \\ 
2 & Generate & 0.42 \\ 
3 & Creativity & 0.31 \\ 
4 & Solving & 0.28 \\ 
5 & Addressing & 0.25 \\ 
6 & Decision Making & 0.22 \\ 
7 & Accountability & 0.22 \\ 
8 & Teamwork & 0.21 \\ 
9 & Critical Thinking & 0.19 \\ 
10 & Sharing & 0.18 \\ 
   \hline
\hline
\end{longtable}

\begin{longtable}{llr}
\caption{Top Ten Soft Skills for Programs with High-Quality Accreditation}\\
  \hline
Rank & Soft Skill & Eigen.vector \\ 
  \hline
1 & Create & 0.85 \\ 
2 & Solving & 0.32 \\ 
3 & Accountability & 0.25 \\ 
4 & Generate & 0.24 \\ 
5 & Addressing & 0.21 \\ 
6 & Decision Making & 0.19 \\ 
7 & Sharing & 0.19 \\ 
8 & Creativity & 0.18 \\ 
9 & Teamwork & 0.17 \\ 
10 & Management & 0.12 \\ 
   \hline
\hline
\end{longtable}

\newpage
\begin{longtable}{rlr}
\caption{Top Ten Soft Skills for Programs with Qualified Certification}\\
  \hline
Rank & Soft Skill & Eigen.vector \\ 
  \hline
1 & Generate & 0.90 \\ 
2  & Creativity & 0.87 \\ 
3  & Identify & 0.42 \\ 
4  & Leadership & 0.33 \\ 
5  & Evaluate & 0.31 \\ 
6  & Acknowledge & 0.31 \\ 
7  & Analytical & 0.28 \\ 
8  & Teamwork & 0.26 \\ 
9  & Create & 0.25 \\ 
10 & Innovate & 0.25 \\ 
   \hline
\hline
\end{longtable}


\begin{thebibliography}{10}

\bibitem{Jamali2023}
S.~M. Jamali, N.~Ale~Ebrahim, and F.~Jamali.
\newblock {The role of STEM Education in improving the quality of education: a
  bibliometric study}.
\newblock {\em International Journal of Technology and Design Education},
  33(3):819--840, 2023.

\bibitem{Harvey2022}
Tracy Harvey, Alfonso Morales, and Catherine~H. Middlecamp.
\newblock Defining sustainability in higher education institutions.
\newblock {\em Sustainability and climate change}, 15(3):182 – 188, 2022.

\bibitem{brauer2021}
Sanna Brauer.
\newblock Towards competence-oriented higher education: a systematic literature
  review of the different perspectives on successful exit profiles.
\newblock {\em Education+ Training}, 63(9):1376--1390, 2021.

\bibitem{dell'aquila2016}
Elena Dell'Aquila, Davide Marocco, Michela Ponticorvo, Andrea Di~Ferdinando,
  Massimiliano Schembri, and Orazio Miglino.
\newblock {\em Educational games for soft-skills training in digital
  environments: New perspectives}.
\newblock Springer, 2016.

\bibitem{Scheerens2020}
J.~Scheerens, Greetje van~der Werf, and Hester de~Boer.
\newblock {\em Soft Skills in Education: Putting the evidence in perspective}.
\newblock Springer, Cham, 2020.

\bibitem{Awang-Hashim2022635}
Rosna Awang-Hashim, Amrita Kaur, Norhafezah Yusof, S~Kanageswari a/p~Suppiah
  Shanmugam, Nor Aziah~Abdul Manaf, Ainol~Madziah Zubairi, Angelina Yee~Seow
  Voon, and Marzura~Abdul Malek.
\newblock {Reflective and integrative learning and the role of instructors and
  institutions—evidence from Malaysia}.
\newblock {\em Higher Education}, pages 1--20, 2022.

\bibitem{Stark2018}
Stark, P.~B. (2018).
\newblock Before reproducibility must come preproducibility.
\newblock {\em Nature}, 557(7706):613--614.

\bibitem{succi2019}
Chiara Succi.
\newblock {Are you ready to find a job? Ranking of a list of soft skills to
  enhance graduates' employability}.
\newblock {\em International Journal of Human Resources Development and
  Management}, 19(3):281--297, 2019.

\bibitem{Succi2020}
Chiara Succi and Magali Canovi.
\newblock {Soft skills to enhance graduate employability: comparing students
  and employers’ perceptions}.
\newblock {\em Studies in Higher Education}, 45(9):1834--1847, 2020.

\bibitem{Coelho202278}
M.J. Coelho and H.~Martins.
\newblock The future of soft skills development: a systematic review of the
  literature of the digital training practices for soft skills.
\newblock {\em Journal of E-Learning and Knowledge Society}, 18(2):78--85,
  2022.

\bibitem{Dolce2020}
Valentina Dolce, Federica Emanuel, Maurizio Cisi, and Chiara Ghislieri.
\newblock {The soft skills of accounting graduates: Perceptions versus
  expectations}.
\newblock {\em Accounting Education}, 29(1):57--76, 2020.

\bibitem{Börner201812630}
K.~Börner, O.~Scrivner, M.~Gallant, S.~Ma, X.~Liu, K.~Chewning, L.~Wu, and
  J.A. Evans.
\newblock Skill discrepancies between research, education, and jobs reveal the
  critical need to supply soft skills for the data economy.
\newblock {\em Proceedings of the National Academy of Sciences of the United
  States of America}, 115(50):12630--12637, 2018.

\bibitem{Rovida20231541}
E.G.M. Rovida, A.~Gianotti, and G.~Zafferri.
\newblock Soft skills teaching proposal for “designers”.
\newblock {\em Lecture Notes in Mechanical Engineering}, pages 1541--1551,
  2023.

\bibitem{Riskiyana20222174}
R.~Riskiyana, N.~Qomariyah, R.N. Hidayah, and M.~Claramita.
\newblock Towards improving soft skills of medical education in the 21st
  century: A literature review.
\newblock {\em International Journal of Evaluation and Research in Education},
  11(4):2174--2181, 2022.

\bibitem{Hamid2022263}
A.~Hamid and M.~Younus.
\newblock Why soft skills matter: Analyzing the relationship between soft
  skills and productivity in workplace of academic library professionals.
\newblock {\em Libri}, 72(3):263--277, 2022.

\bibitem{Medvedeva2022}
O.D. Medvedeva, A.V. Rubtsova, A.V. Vilkova, and V.V. Ischenko.
\newblock Digital monitoring of students’ soft skills development as an
  interactive method of foreign language learning.
\newblock {\em Education Sciences}, 12(8), 2022.

\bibitem{Daniels202390}
R.A. Daniels, S.D. Pemble, D.~Allen, G.~Lain, and L.A. Miller.
\newblock Linkedin blunders: A mixed method study of college students’
  profiles.
\newblock {\em Community College Journal of Research and Practice},
  47(2):90--105, 2023.

\bibitem{Healy2023106}
M.~Healy, S.~Cochrane, P.~Grant, and M.~Basson.
\newblock Linkedin as a pedagogical tool for careers and employability
  learning: a scoping review of the literature.
\newblock {\em Education and Training}, 65(1):106--125, 2023.

\bibitem{Alvarez2022}
M.J. Alvarez-Rivadulla, A.M. Jaramillo, F.~Fajardo, L.~Cely, A.~Molano, and
  F.~Montes.
\newblock College integration and social class.
\newblock {\em Higher Education}, 84(3):647--669, 2022.

\bibitem{Duque2021669}
J.F. Duque.
\newblock {A comparative analysis of the Chilean and Colombian systems of
  quality assurance in higher education}.
\newblock {\em Higher Education}, 82(3):669--683, 2021.

\bibitem{Bradford2018909}
H.~Bradford, A.~Guzmán, J.M. Restrepo, and M.-A. Trujillo.
\newblock {Who controls the board in non-profit organizations? The case of
  private higher education institutions in Colombia}.
\newblock {\em Higher Education}, 75(5):909--924, 2018.

\bibitem{Berry2014}
C.~Berry and J.~Taylor.
\newblock {Internationalisation in higher education in Latin America: Policies
  and practice in Colombia and Mexico}.
\newblock {\em Higher Education}, 67(5):585--601, 2014.

\bibitem{Melguizo2011}
T.~Melguizo, F.S. Torres, and H.~Jaime.
\newblock {The association between financial aid availability and the college
  dropout rates in Colombia}.
\newblock {\em Higher Education}, 62(2):231--247, 2011.

\bibitem{Jaimes2022}
Y.-C. Jaimes-Acero, A.~Granados-Comba, and R.~Bolivar-Leon.
\newblock {Soft Skills Requirements for Engineering Entrepreneurship}.
\newblock {\em {Revista Facultad de Ingeniería}}, 31, 03 2022.

\bibitem{Renteria2022}
J.~A. Rentería-Vera, E.~M. Hincapi-Montoya, Y.~J. Rodríguez-Caro, C.~K.
  Vélez-Castaneda, B.~E. Osorio-Vélez, and J.~A. Durango-Marín.
\newblock {Competencia global para el desarrollo sostenible: una oportunidad
  para la educación superior}.
\newblock {\em {Entramado}}, 18:e208, 06 2022.

\bibitem{Ocde2016-nq}
OECD.
\newblock {\em Perspectivas económicas de América Latina 2017}.
\newblock OECD Publishing, 2016.

\bibitem{Zarate2023}
R.~Zarate-Torres and J.~C. Correa.
\newblock {How Good is the Myers-Briggs Type Indicator for Predicting
  Leadership-Related Behaviors?}
\newblock {\em Frontiers in Psychology}, 14:14:940961, 2023.

\bibitem{Dellavigna2010}
Stefano DellaVigna and Matthew Gentzkow.
\newblock Persuasion: empirical evidence.
\newblock {\em Annual Review of Economics}, 2(1):643--669, 2010.

\bibitem{Lorange2019}
P.~Lorange.
\newblock {\em The Business School of the Future}.
\newblock Cambridge University Press, 2019.

\bibitem{Estrada2011}
E.~Estrada.
\newblock {\em {The Structure of Complex Networks: Theory and Applications}}.
\newblock Oxford University Press, 2011.

\bibitem{Glass2023}
C.~R. Glass and N.~I. Cruz.
\newblock Moving towards multipolarity: shifts in the core‑periphery
  structure of international student mobility and world rankings (2000–2019).
\newblock {\em Higher Education}, 85:415--435, 2023.

\bibitem{Gandrud2018}
Christopher Gandrud.
\newblock {\em Reproducible research with R and RStudio}.
\newblock Chapman and Hall/CRC, 2018.

\bibitem{bisquerra2007}
Rafael Bisquerra-Alzina and Nuria Escoda-P{\'e}rez.
\newblock Las competencias emocionales.
\newblock {\em Educación XX1}, 10:61--82, 2007.

\bibitem{vera2020}
Fernando Vera and Eneko Tejada.
\newblock {Developing soft skills in undergraduate students: A case at a
  Chilean private university}.
\newblock {\em Transformar}, 1(1):57--67, 2020.

\bibitem{Botke2018}
J.A. Botke, P.G.W. Jansen, S.N. Khapova, and M.~Tims.
\newblock Work factors influencing the transfer stages of soft skills training:
  A literature review.
\newblock {\em Educational Research Review}, 24:130--147, 2018.

\bibitem{goleman1998}
Daniel Goleman.
\newblock {\em Working with emotional intelligence}.
\newblock Bantam, 1998.

\bibitem{zins2004}
J.~E. Zins, R.~P. Weissberg, M.~C. Wang, and H.~J. Walberg.
\newblock {\em {Building academic success on social and emotional learning:
  What does the research say?}}
\newblock Teachers College Press, New York, 2004.

\bibitem{abdullah2012}
Abdul Ghani~Kanesan Abdullah, Sim~Hock Keat, Aziah Ismail, Mohamad~Hanif
  Abdullah, and Miduk Purba.
\newblock {Mismatch between higher education and employment in Malaysian
  electronic industry: an analysis of the acquired and required competencies}.
\newblock {\em The International journal of engineering education},
  28(5):1232--1242, 2012.

\bibitem{sharvari2019}
Kulkarni Sharvari and DG~Kulkarni.
\newblock {Gap analysis of soft skills in the curriculum of higher education (A
  case study of management institutes in Karnataka)}.
\newblock {\em Advances in Management}, 12(1):64--67, 2019.

\bibitem{volkova2020}
Nataliia Volkova, Nataliia Zinukova, Kateryna Vlasenko, and Tetiana
  Korobeinikova.
\newblock Soft skills, their development and mastering among post graduate
  students.
\newblock {\em SHS Web of Conferences}, 75:04002, 2020.

\bibitem{ziberi2021}
Besime Ziberi, Donat Rexha, and Kosovare Ukshini.
\newblock {Skills mismatch in the labor market: The future of work from the
  viewpoint of enterprises in case of Kosovo}.
\newblock {\em Journal of Governance and Regulation/Volume}, 10(3), 2021.

\bibitem{Cattani2021}
Luca Cattani and Giulio Pedrini.
\newblock {Opening the black-box of graduates’ horizontal skills: diverging
  labour market outcomes in Italy}.
\newblock {\em Studies in Higher Education}, 46(11):2387--2404, 2021.

\bibitem{cieciora2021}
Ma{\l}gorzata Cieciora, Piotr Pietrzak, and Piotr Gago.
\newblock {University graduates' skills-and-employability evaluation in Poland
  --a case study of a faculty of management in Warsaw}.
\newblock {\em International Journal of Innovation and Learning}, 30(1):1--18,
  2021.

\bibitem{Benlahcene2022}
A.~Benlahcene, O.~Saoula, M.~Jaganathan, A.~Ramdani, and N.A. AlQershi.
\newblock The dark side of leadership: How ineffective training and poor ethics
  education trigger unethical behavior?
\newblock {\em Frontiers in Psychology}, 13, 2022.

\bibitem{Joie2023}
Chantal Joie-La~Marle, Fran{\c{c}}ois Parmentier, Pierre-Louis Weiss, Martin
  Storme, Todd Lubart, and Xavier Borteyrou.
\newblock {Effects of a New Soft Skills Metacognition Training Program on
  Self-Efficacy and Adaptive Performance}.
\newblock {\em Behavioral Sciences}, 13(3):202, 2023.

\bibitem{Kuckertz2023}
Andreas Kuckertz, Maximilian Scheu, and Per Davidsson.
\newblock {Chasing mythical creatures--A (not-so-sympathetic) critique of
  entrepreneurship's obsession with unicorn startups}.
\newblock {\em Journal of Business Venturing Insights}, 19:e00365, 2023.

\bibitem{CNA2008}
{Consejo Nacional de Acreditación}.
\newblock {\em {Lineamientos para la Acreditación de Alta Calidad de Programas
  de Maestría y Doctorado}}.
\newblock Ministerio de Educación, Bogotá, Colombia, 2008.

\bibitem{Correa2020}
J.~C. Correa.
\newblock {Metrics of Emergence, Self-Organization, and Complexity for EWOM
  Research}.
\newblock {\em Frontiers in Physics}, 8:35, 2020.

\bibitem{Luke2015}
D.~A. Luke.
\newblock {\em {A User’s Guide to Network Analysis in R}}.
\newblock Springer, Cham, 2015.

\bibitem{Oldham2019}
Stuart Oldham, Ben Fulcher, Linden Parkes, Aurina Arnatkevici{\=u}t{\.e},
  Chao Suo, and Alex Fornito.
\newblock Consistency and differences between centrality measures across
  distinct classes of networks.
\newblock {\em PloS one}, 14(7):e0220061, 2019.

\bibitem{ronqui2015}
Jos{\'e} Ricardo~Furlan Ronqui and Gonzalo Travieso.
\newblock Analyzing complex networks through correlations in centrality
  measurements.
\newblock {\em Journal of Statistical Mechanics: Theory and Experiment},
  2015(5):P05030, 2015.

\bibitem{Saura2022}
J.R. Saura, D.~Ribeiro-Soriano, and D.~Palacios-Marqués.
\newblock Assessing behavioral data science privacy issues in government
  artificial intelligence deployment.
\newblock {\em Government Information Quarterly}, 39(4), 2022.

\bibitem{Manning2008}
C.~D. Manning, P.~Raghavan, and H.~Schütze.
\newblock {\em Introduction to Information Retrieval}.
\newblock Cambridge Univesity Press, 2008.

\bibitem{RCore2022}
{R Core Team}.
\newblock {\em R: A Language and Environment for Statistical Computing}.
\newblock R Foundation for Statistical Computing, Vienna, Austria, 2023.

\bibitem{Wickham2019}
H.~Wickham and G.~Grolemund.
\newblock {\em {R for Data Science}}.
\newblock O'Reilly, California, USA, 2017.

\bibitem{Feinerer2008}
Ingo Feinerer, Kurt Hornik, and David Meyer.
\newblock {Text Mining Infrastructure in R}.
\newblock {\em Journal of Statistical Software}, 25(5):1--54, 2008.

\bibitem{Benoit2018}
Kenneth Benoit, Kohei Watanabe, Haiyan Wang, Paul Nulty, Adam Obeng, Stefan
  Müller, and Akitaka Matsuo.
\newblock {quanteda: An R package for the quantitative analysis of textual
  data}.
\newblock {\em Journal of Open Source Software}, 3(30):774, 2018.

\bibitem{Bail2016}
C.A. Bail.
\newblock Combining natural language processing and network analysis to examine
  how advocacy organizations stimulate conversation on social media.
\newblock {\em Proceedings of the National Academy of Sciences of the United
  States of America}, 113(42):11823--11828, 2016.

\bibitem{Fletcher2023}
S.~Fletcher and K.R.V. Thornton.
\newblock The top 10 soft skills in business today compared to 2012.
\newblock {\em Business and Professional Communication Quarterly}, 2023.

\bibitem{Marin2022}
S.I. Marin-Zapata, J.P. Román-Calderón, C.~Robledo-Ardila, and M.A.
  Jaramillo-Serna.
\newblock Soft skills, do we know what we are talking about?
\newblock {\em Review of Managerial Science}, 16(4):969--1000, 2022.

\bibitem{Jang2016}
Hyewon Jang.
\newblock Identifying 21st century stem competencies using workplace data.
\newblock {\em Journal of science education and technology}, 25:284--301, 2016.

\bibitem{Hassan2023511}
S.~Hassan, P.~Kaur, M.~Muchiri, C.~Ogbonnaya, and A.~Dhir.
\newblock Unethical leadership: Review, synthesis and directions for future
  research.
\newblock {\em Journal of Business Ethics}, 183(2):511--550, 2023.

\bibitem{Crawford20231}
J.~Crawford.
\newblock Editorial: The need for good leaders in higher education.
\newblock {\em Journal of University Teaching and Learning Practice},
  20(1):1--7, 2023.

\bibitem{hyytinen2023}
Heidi Hyytinen, Kari Nissinen, Katri Kleemol, Jani Ursin, and Auli Toom.
\newblock How do self-regulation and effort in test-taking contribute to
  undergraduate students’ critical thinking performance?
\newblock {\em Studies in Higher Education}, pages 1--14, 2023.

\bibitem{Campo2023}
Luc{\'\i}a Campo, H{\'e}ctor Galindo-Dom{\'\i}nguez, Mar{\'\i}a-Jos{\'e}
  Bezanilla, Donna Fern{\'a}ndez-Nogueira, and Manuel Poblete.
\newblock Methodologies for fostering critical thinking skills from university
  students’ points of view.
\newblock {\em Education Sciences}, 13(2):132, 2023.

\bibitem{LiCausi2022}
T.J. LiCausi and D.A. McFarland.
\newblock Abstract(s) at the core: a case study of disciplinary identity in the
  field of linguistics.
\newblock {\em Higher Education}, 84(5):955--978, 2022.

\bibitem{Mantai2022}
L.~Mantai and M.~Marrone.
\newblock {Identifying skills, qualifications, and attributes expected to do a
  PhD}.
\newblock {\em Studies in Higher Education}, 47(11):2273--2286, 2022.

\bibitem{Suyansah2023}
Q.~Suyansah, D.~Gabda, E.~Jawing, K.~Kamlun, R.P. Tibok, H.~Wendy, and W.N.Y.
  Xe.
\newblock {Students' academic performance and soft skills on graduate
  employability among students in Universiti Malaysia Sabah}.
\newblock {\em AIP Conference Proceedings}, 2500:020038, 2023.

\bibitem{Feraco2023}
T.~Feraco, D.~Resnati, D.~Fregonese, A.~Spoto, and C.~Meneghetti.
\newblock An integrated model of school students’ academic achievement and
  life satisfaction. linking soft skills, extracurricular activities,
  self-regulated learning, motivation, and emotions.
\newblock {\em European Journal of Psychology of Education}, 38(1):109--130,
  2023.

\bibitem{van2013}
Rens Van~de Schoot, Mara~A Yerkes, Jolien~M Mouw, and Hans Sonneveld.
\newblock {What took them so long? Explaining PhD delays among doctoral
  candidates}.
\newblock {\em PloS one}, 8(7):e68839, 2013.

\end{thebibliography}
\end{document}